\documentclass[fleqn,usenatbib]{mnras}

\usepackage{newtxtext,newtxmath}
\usepackage{amsmath}
\usepackage{graphicx}

\usepackage[T1]{fontenc}
\usepackage{mathtools}

\newcommand{\vsh}{\ensuremath{v_\text{sh}}}
\newcommand{\msh}{\ensuremath{m_\text{sh}}}
\newcommand{\rsh}{\ensuremath{R_\text{sh}}}
\newcommand{\Lshock}{\ensuremath{L_\text{sh}}}

\newcommand{\rhocsm}{\ensuremath{\rho_\text{CSM}}}
\newcommand{\rhosn}{\ensuremath{\rho_\text{SN}}}
\newcommand{\mdot}{\ensuremath{\dot{M}}}
\newcommand{\vesc}{\ensuremath{v_\text{esc}}}
\newcommand{\teff}{\ensuremath{T_{\text{eff}}}}
\newcommand{\mcw}{\ensuremath{M_\text{cw}}}
\newcommand{\vtan}{\ensuremath{v_\text{tan}}}
\newcommand{\msun}{\ensuremath{\rm M_\odot}}
\newcommand{\rsun}{\ensuremath{R_\odot}}
\newcommand{\myr}{\ensuremath{\msun\,\text{yr}^{-1}}}
\newcommand{\kms}{\ensuremath{\text{km}\,\text{s}^{-1}}}

\usepackage{amsmath}
\title[Supernovae in colliding-wind binaries]{Supernovae in colliding-wind binaries: observational signatures in the first year}

\author[]{Ond\v{r}ej Pejcha,$^{1}$\thanks{E-mail: pejcha@utf.mff.cuni.cz}
Diego Calder\'{o}n,$^{1}$
Petr Kurf\"{u}rst$^{2}$
\\
$^{1}$Institute of Theoretical Physics, Faculty of Mathematics and Physics, Charles University, V Hole\v{s}ovi\v{c}k\'{a}ch 2, Prague 8, 18000, Czech Republic\\
$^{2}$Department of Theoretical Physics and Astrophysics, Masaryk University, Kotlářská 2, 611 37 Brno, Czech Republic
}

\date{Accepted XXX. Received YYY; in original form ZZZ}

\pubyear{2021}

\begin{document}
\label{firstpage}
\pagerange{\pageref{firstpage}--\pageref{lastpage}}
\maketitle

\begin{abstract}
When a core-collapse supernova explodes in a binary star system, the ejecta might encounter an overdense shell, where the stellar winds of the two stars previously collided. In this work, we investigate effects of such interactions on supernova light curves on time-scales from the early flash ionization signatures to approximately one year after the explosion. We construct a model of the colliding-wind shell in an orbiting binary star system and we provide an analytical expression for the shell thickness and density, which we calibrate with three-dimensional adaptive mesh refinement hydrodynamical simulations probing different ratios of wind momenta and different regimes of radiative cooling efficiency. We model the angle-dependent interaction of supernova ejecta with the circumstellar medium and estimate the shock radiative efficiency with a realistic cooling function. We find that the radiated shock power exceeds typical Type IIP supernova luminosity only for double red supergiant binaries with mass ratios $q \gtrsim 0.9$, wind mass-loss rates $\mdot \gtrsim 10^{-4}\,\myr$, and separations between about $50$ and $1500$\,au. The required $\mdot$ increases for binaries with smaller $q$ or primaries with faster wind. We estimate that $\ll 1$~per~cent of all collapsing massive stars satisfy the conditions on binary mass ratio and separation.  Recombination luminosities due to colliding wind shells are at most a factor of $10$ higher than for an otherwise unperturbed constant-velocity wind, but higher densities associated with wind acceleration close to the star provide much stronger signal. 
\end{abstract}

\begin{keywords}
binaries:general -- stars:massive -- supernovae: general -- stars: winds, outflows
\end{keywords}



\section{Introduction}

The first electromagnetic signature of a core-collapse supernova (SN) should be a break-out pulse of UV and X-ray radiation, which is caused by the SN shock approaching the progenitor surface. This bright burst of radiation ionizes the circumstellar medium (CSM), which then recombines on a time-scale of days. Signatures of this process can be observed with ``flash spectroscopy'' soon after the SN explosion \citep[e.g.][]{galyam14,khazov16,yaron17,groh14,bruch21}. When the SN ejecta physically collide with the CSM, an interaction region forms between the forward shock propagating into the CSM and the reverse shock travelling back inside the SN ejecta. High densities and temperatures between the two shocks can make the radiative cooling time shorter than the expansion time and, as a consequence, the slab between the two shocks collapses to a thin shell. Radiative cooling of the shocked region can make the SN considerably more luminous as well as change its spectroscopic appearance \citep[e.g.][]{chugai94,chugai04,dessart15,smith17}. 

Observations of SNe have revealed a puzzling diversity of CSM surrounding the progenitor star at distances of 10s to 1000s of au. CSM properties combined with estimates of shell expansion velocity based on the progenitor properties imply that many massive stars lose substantial amount of mass shortly before their core collapses \citep[e.g.][]{smith07,smith14,moriya13}. This connection is verified by direct records of pre-SN outbursts in the progenitors of Type II-n SNe  \citep[e.g.][]{mauerhan13,ofek14}. Flash ionization observations suggest than more than $30$ per cent of hydrogen-rich SNe have CSM overdensities located near the progenitor. Shock power from SN--CSM collisions is also used to explain early light curves of ordinary Type II-P SNe, but their progenitors seem to be relatively quiet with low levels of pre-SN variability \citep[e.g.][]{morozova17,morozova18,johnson18,jacobsongalan21}. Adding to the puzzle, spectral line profiles and (spectro)polarimetry suggest that the CSM around many SNe lacks spherical symmetry \citep[e.g.][]{chugai94,leonard00,andrews18}.

The origin of the CSM overdensities remains mysterious and a number of theories have been proposed to explain the observations. For example, turbulent nuclear burning in the final evolution stages of the progenitor can lead to wave-driven mass loss shortly before the SN explosion \citep[e.g.][]{quataert12,smith_arnett14,wu21}. Another example are strong interactions in binary star systems such as the common envelope ejection, which can precede a SN explosion by a relatively short time. In this case, the CSM geometry will reflect the orbital plane symmetry of the binary star \citep[e.g.][]{podsiadlowski92,morris07,pejcha16}. 

Alternatively, smooth and steady wind from the progenitor can be compressed by external effects such as ambient ionizing photons or a wind from a stellar companion. For example, \citet{mackey14} showed that red supergiant winds will form an overdense shell due to ambient photoionizing radiation and a more distant bow shock due to the collision with the interstellar medium. \citet{ryder04} interpreted quasi-periodic oscillations in the radio light curve of Type II-b SN 2001ig as a signature of spiral overdensity formed by colliding winds (CW) of two Wolf-Rayet stars. \citet{kochanek19} argued that flash ionization signatures can be dominated by CW shells and proposed this as an explanation for SN 2013fs, where the early spectroscopy of \citet{yaron17} reveals H$\alpha$ brightening starting few days after the explosion. Motivated by these works, \citet{kurfurst20} performed hydrodynamical simulations of a spherical SN explosion colliding with a thin bow-shock shell with various orientations. They characterized the development of various hydrodynamical instabilities and estimated time evolution of shock power, spectroscopic line profiles from various viewing angles, and polarization signatures. They concluded that aspherical shells have the potential to provide high shock-interaction luminosities and might explain some very puzzling observations such as asymmetric spectral line profiles with evolving blue and red wings \citep[e.g.][]{smith15,andrews17,bilinski18,bilinski20}. However, shell properties in \citet{kurfurst20} were manually chosen to get a strong hydrodynamic effect and might not be entirely realistic.

CW shells are an attractive possibility for explaining the CSM overdensities, because predicting their structure based on fundamental properties of the binary system is relatively straightforward, unlike intrinsic mechanisms of mass ejection. However, this does not mean that characterising the detailed structure and physical properties of wind-confined slabs is not challenging. Pioneering work by \cite{stevens92} provided the theoretical basis for describing the location and the radiative properties of CW shells. \cite{canto96} derived analytically expressions for the location and shape of the CW shell given the properties of the winds. Unfortunately, there is no analytical formalism for calculating the hydrodynamic and thermodynamic properties, especially if the shell is subject to hydrodynamic and/or thermal instabilities \citep{vishniac83,vishniac94}. Despite the existence of many sophisticated models of CW binaries \citep[e.g.][]{pittard09,parkin11,kee14,hendrix16}, only a few have focused on making quantitative characterizations of the (unstable) shells. 
\cite{lamberts11,lamberts12} conducted the first high-resolution simulations of wind collisions aiming to characterise the mechanisms responsible of the instabilities and the impact of orbital motion. 
\cite{vanmarle11} performed three-dimensional simulations of unstable wind collisions in both radiatively efficient and inefficient regimes combined with the orbital motion of the binary. 
\citet{steinberg18} studied corrugated radiative shocks in the context of X-ray emission and high-energy particle acceleration. 
A more detailed characterisation of the substructures formed in unstable wind collisions was done by \cite{calderon20b}, who utilized three-dimensional adaptive-mesh refinement (AMR) simulations to resolve the thin slabs and obtain their properties. 
With this capability at hand, it is now possible to get physically motivated slab structures as CSM models in the context of SN light curve calculations.

The goal of this paper is to develop a realistic semi-analytic model of CW shells in various settings and to assess the observability of these shells in SN light curves powered by shock interaction and in spectroscopically-observed recombination light curves. In Section~\ref{sec:csm}, we modify an existing semi-analytic model of density structure of colliding-winds of \citet{canto96} and calibrate it in various regimes with three-dimensional AMR hydrodynamical simulations. In Section~\ref{sec:lc}, we present a model of the angle-dependent dynamics of the shocked shell. In Section~\ref{sec:results}, we discuss the implications for optical light curves, spectroscopic recombination light curves, and we estimate the event rates. In Section~\ref{sec:disc}, we summarize and discuss our findings.

\section{Circumstellar medium in a binary star with a colliding wind shell}
\label{sec:csm}

Position and shape of a thin CW shell was derived analytically by \citet{canto96} by considering the balance of momentum at the interface of two isotropic stellar winds. Here, we review the formalism of \citet{canto96} in Section~\ref{sec:thin_shell} and address two main uncertainties in obtaining a realistic density distribution. First, the thickness of the shell is unconstrained by the analytic model and depends on the efficiency of the radiative cooling as well as on the presence or absence of instabilities. In Section~\ref{sec:thick}, we present a prescription for shell thickness that represents mean properties of three-dimensional hydrodynamic simulations in both stable adiabatic and unstable radiatively-efficient regimes. Second, \citet{canto96} model does not take into account orbital motion of the two stars, which becomes especially important when the wind speeds are similar to the orbital velocity. In Section~\ref{sec:density}, we describe a heuristic model of a smooth transition between the CW shell and the outer regions where the two winds are completely mixed. We conclude by giving the final expression for the density distribution.

\subsection{Analytic model of a thin shell}
\label{sec:thin_shell}

\begin{figure}
    \centering
    \includegraphics[width=0.48\textwidth]{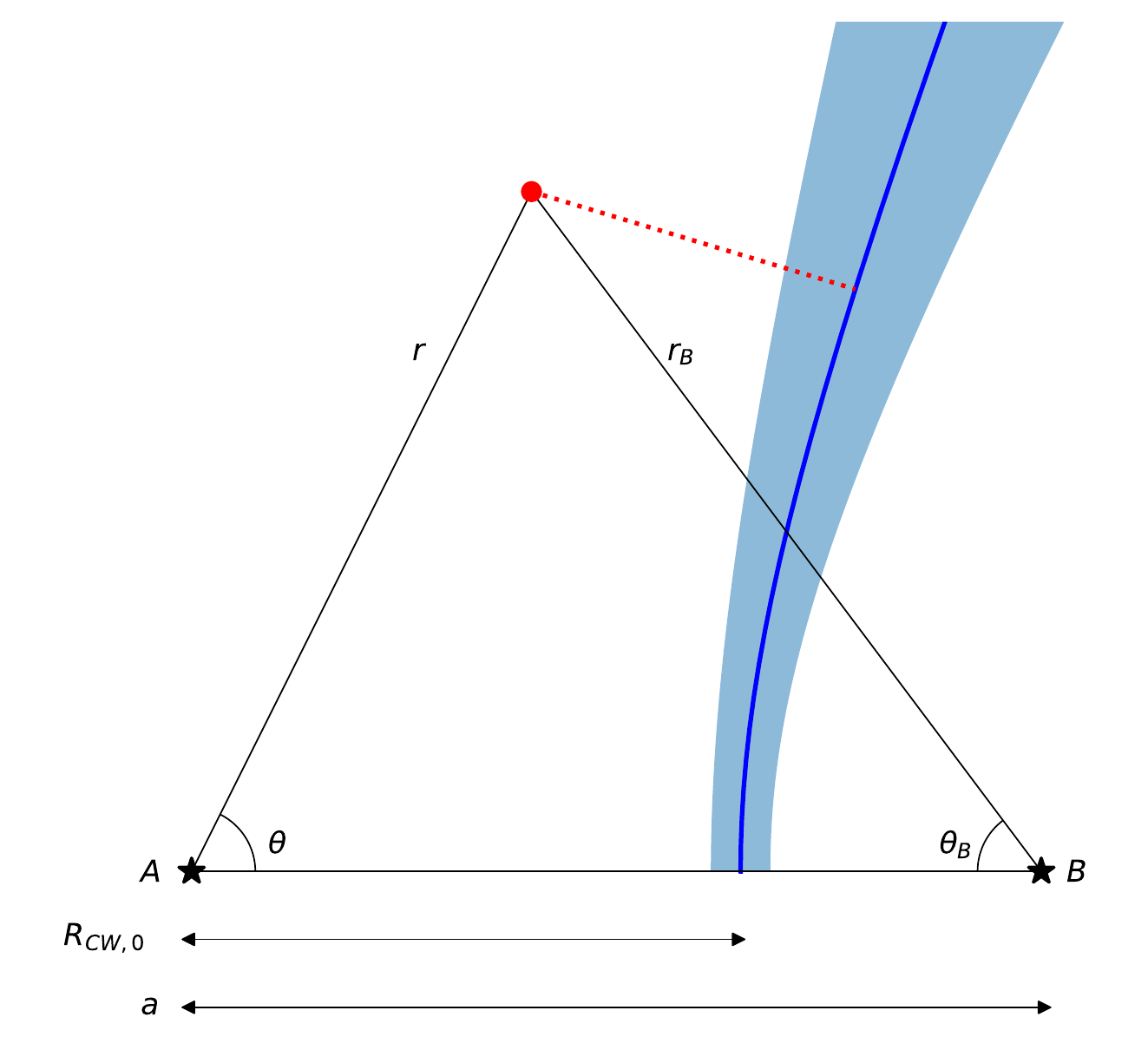}
    \caption{Diagram of our coordinate system. Red point with coordinates $(r,\theta)$ as seen from star $A$ has distance $r_B$ and angle $\theta_B$ from star $B$. The center of the colliding wind shell is shown with solid blue line and its width is indicated by the light blue shaded region. The dotted red line marks the nearest point on the colliding wind shell to the point at $(r,\theta)$.}
    \label{fig:angle}
\end{figure}

In Figure~\ref{fig:angle}, we show the schematic diagram of the colliding wind shell. We place two stars with masses $M_A$ and $M_B$, and radii $R_A$ and $R_B$ at a mutual distance $a$. We assume that each star has a time-steady isotropic wind with density given by
\begin{equation}
    \rho_{A}(r) = \frac{\mdot_{A}}{4\pi r^2 v_{\infty,A}},\quad \rho_{B}(r_B) = \frac{\mdot_{B}}{4\pi r_B^2 v_{\infty,B}},
    \label{eq:rho}
\end{equation}
where $r$ is distance from $A$ and $r_B$ is distance from $B$, $\mdot_A$ and $\mdot_B$ are wind mass-loss rates, and $v_{\infty,A}$ and $v_{\infty,B}$ are wind terminal velocities. It is convenient to define ratios of wind velocities and momenta
\begin{equation}
    \alpha = \frac{v_{\infty,A}}{v_{\infty,B}},\quad \beta = \frac{\mdot_A v_{\infty,A}}{\mdot_B v_{\infty,B}} = \alpha \frac{\mdot_A}{\mdot_B}.
    \label{eq:alphabeta}
\end{equation}
Realistic stellar winds accelerate from slow velocities near the stellar photosphere to the asymptotic velocity far from the star. For many binaries of interest, the radial scale of wind acceleration is comparable to $a$. However, taking wind acceleration into account would significantly complicate the analytic model and its calibration so we leave it aside for now and defer the discussion to Section~\ref{sec:disc}. 

\begin{figure}
    \centering
    \includegraphics[width=0.48\textwidth]{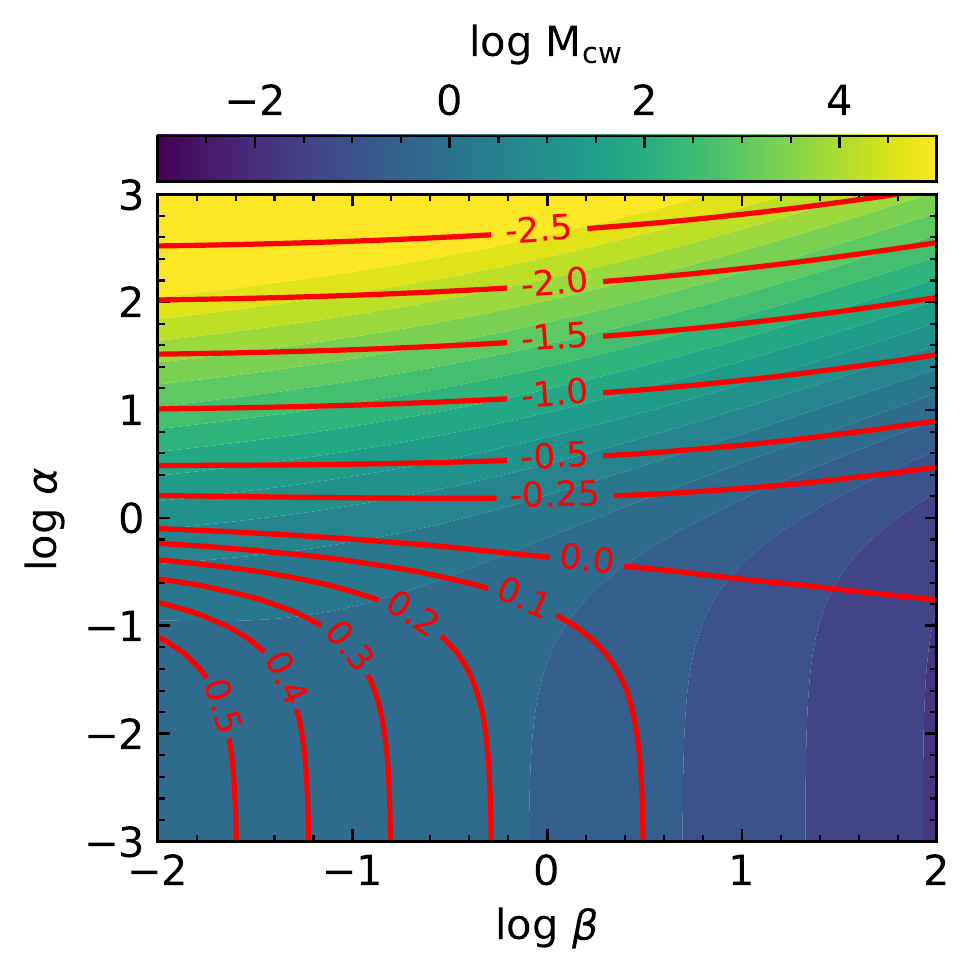}
    \caption{Mass in the CW shell $\mcw$ and the typical tangential velocity $\vtan$ as a function of wind parameters $\alpha$ and $\beta$. CW shell mass is shown in color (blue to yellow) in units of $a\mdot_A/v_{\infty,A}$. Contours of tangential velocity are shown in red with labels giving $\log \vtan$ in the units of $v_{\infty,A}$. Both quantities are evaluated at distance $2a$ from star $A$.}
    \label{fig:canto}
\end{figure}

As shown by \citet{canto96}, CW shell is located at a distance $R_\text{cw}$ from $A$, where
\begin{equation}
    R_\text{cw} = a \sin \theta_B \csc (\theta + \theta_B) = a f(\theta),
    \label{eq:shell_R}
\end{equation}
where $\theta$ is angle between star $B$ and the point on the shell as seen from $A$, while $\theta_B$ is similarly centered on star $B$. Mass and momentum conservation give
\begin{equation}
    \theta_B \cot \theta_B = 1 + \beta(\theta \cot\theta - 1),
    \label{eq:thetas}
\end{equation}
which ties together $\theta$ and $\theta_B$. The stagnation radius on the axis connecting both stars is given by
\begin{equation}
    R_\text{cw,0} = a\frac{{\beta}^{1/2}}{1+{\beta}^{1/2}}.
    \label{eq:r0}
\end{equation}
The asymptotic angle of the CW shell is obtained by solving 
\begin{equation}
    \theta_\infty - \tan \theta_\infty = \frac{\pi}{1-\beta}.
    \label{eq:thetainfty}
\end{equation}
Surface density of the shell is
\begin{equation}
    \sigma = \frac{\mdot_A}{2\pi\beta a v_{\infty,A}} \frac{\mathcal{A}}{\mathcal{B}},
    \label{eq:sigma}
\end{equation}
where the coefficients are
\begin{align}
    \mathcal{A} ={} &\sin(\theta + \theta_B)  \csc\theta \csc\theta_B \times \nonumber \\
    & \times [\beta (1 - \cos\theta) + \alpha (1-\cos\theta_B)]^2, \\
    \mathcal{B}^2 = {} & 
    [
    \beta (\theta - \sin\theta\cos\theta) + \nonumber\\
         & + (\theta_B - \sin\theta_B\cos\theta_B)
    ]
    ^2 + [\beta \sin^2\theta - \sin^2\theta_B]^2.
\end{align}
The tangential velocity of the material in the shell is
\begin{equation}
    \vtan =  \frac{1}{2}\frac{v_{\infty,A}\mathcal{B}}{\beta(1-\cos\theta) + \alpha(1-\cos\theta_B  )}.
    \label{eq:vtan}
\end{equation}
It is interesting to know the total mass contained in the shell $\mcw$ and the typical tangential velocity. Following \citet{kochanek19}, we calculate
\begin{equation}
    \mcw = \int\limits_0^{\theta_{2a}} 2\pi a^2 \sigma f(\theta) \sin\theta \sqrt{f(\theta)^2 + \left(\frac{\partial f(\theta)}{\partial\theta}\right)^2} \text{d}\theta,
    \label{eq:mshell}
\end{equation}
where $\theta_{2a} = \theta|_{R_\text{cw}=2a}$. The upper integration limit was chosen somewhat arbitrary to cover only the immediate vicinity of the binary, because $\mcw$ diverges as $R_\text{cw}$ corresponding to the upper integration limit increases. 

\begin{figure*}
    \centering
    \hspace{1cm}\includegraphics[width=0.43\textwidth]{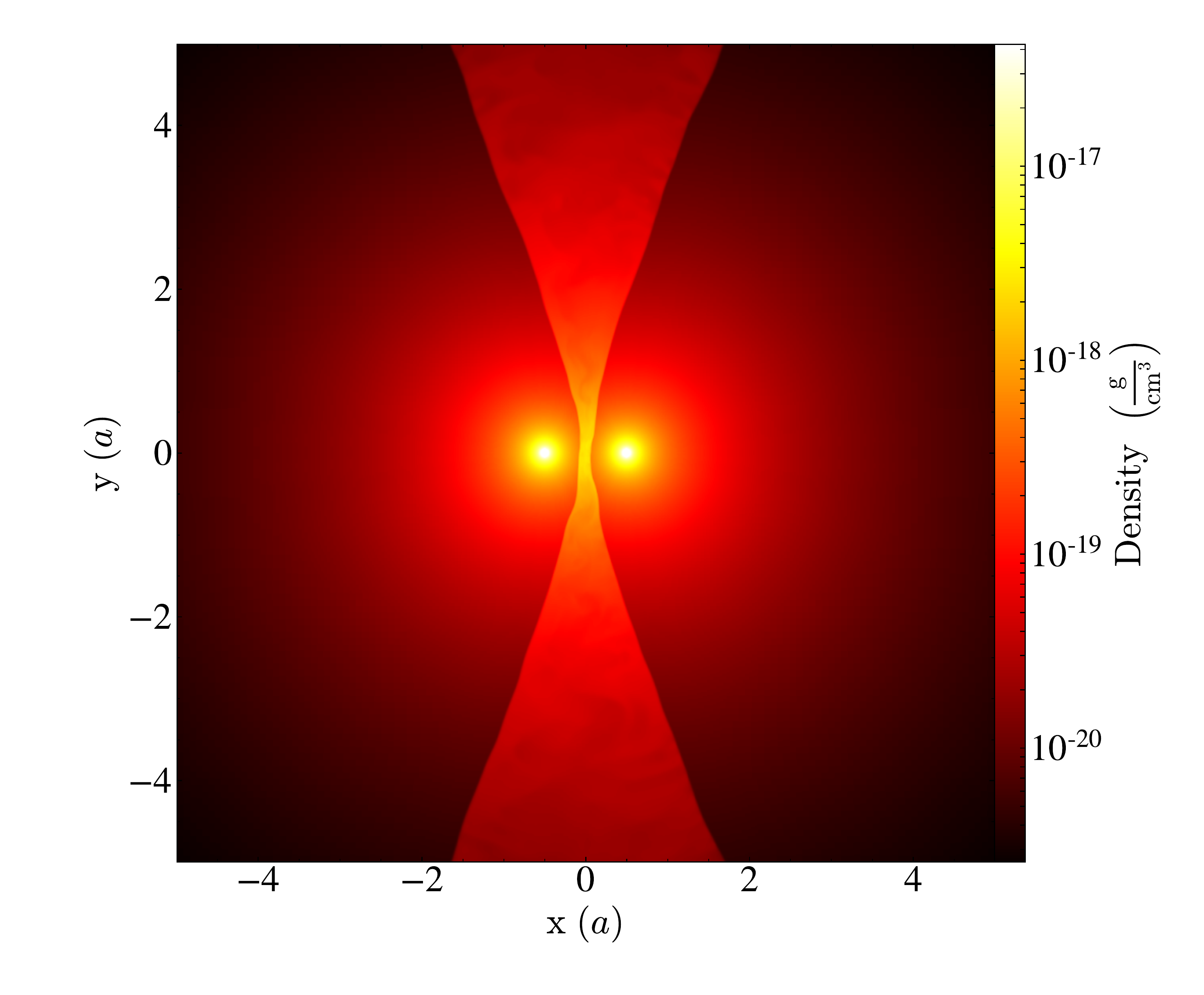}
    \includegraphics[width=0.43\textwidth]{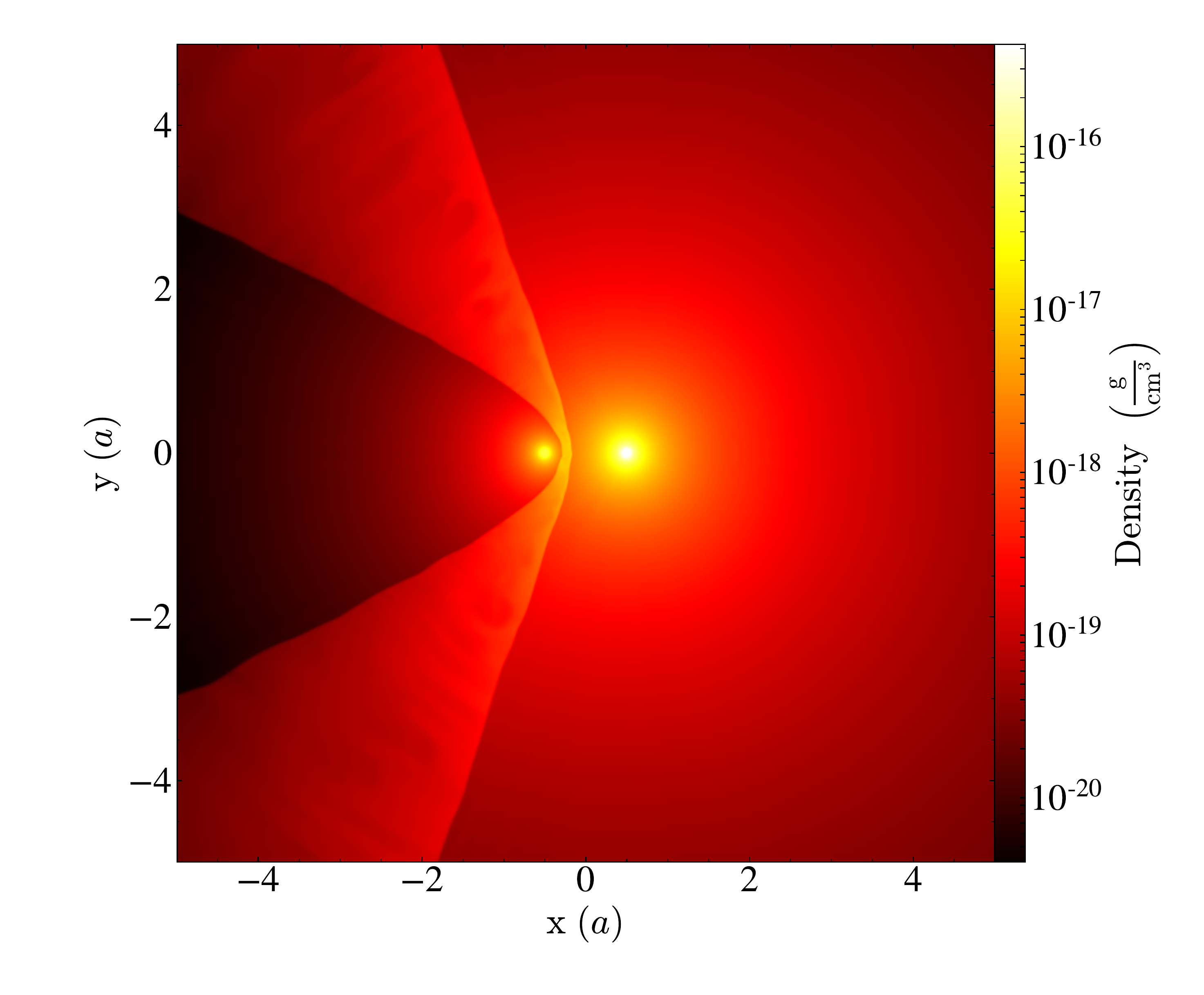}
    \includegraphics[width=0.8\textwidth]{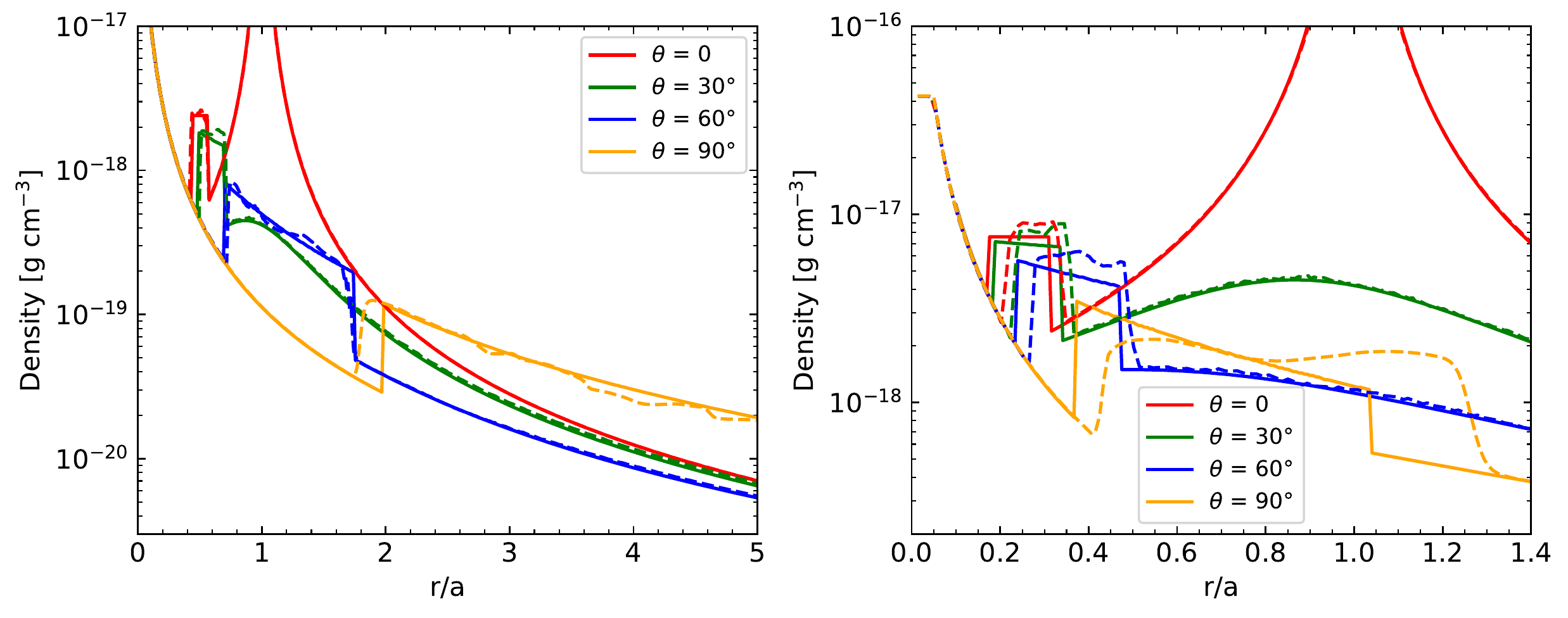}
    \caption{Comparison of hydrodynamical simulations with our analytic model for adiabatic colliding winds. Top row shows density distributions for two identical stars (left) and for unequal wind strengths (right, $\beta = 0.1$, $\alpha=1$). In both plots, star $A$ is on the left side. Bottom row compares density distributions along four rays originating from $A$ with $\theta = 0$ (red), $30\degr$ (green), $60\degr$ (blue), and $90\degr$ (yellow) calculated from hydrodynamical simulations (dashed lines) and our analytic model (solid lines).}
    \label{fig:adiabatic}
\end{figure*}

In Figure~\ref{fig:canto}, we show $\mcw$ and $\vtan(\theta_{2a})$ as a function of $\beta$ and $\alpha$, and taking out the dependence on the physical parameters of the primary star. We see that the relative $\mcw$ and $\vtan$ depend only weakly on the ratio of wind momenta $\beta$, but quite strongly on the ratio of wind velocities $\alpha$. This is caused by the fact that at fixed properties of $A$ and constant $\beta$, $\alpha \propto \mdot_B$ and therefore $\alpha$ is also proportional the total mass flux in the CW shell. As a result, for higher $\alpha$ the CW shell needs to move higher total mass flux, $\mdot_A+\mdot_B$, with the same constant momentum, $\mdot_A v_{\infty,A} + \mdot_B v_{\infty,B} = (1+1/\beta) \mdot_A v_{\infty,A}$. As a result, $\vtan$ is smaller, $\sigma$ is higher, and there is higher $\mcw$. This implies that twin binaries with slow winds such as double red supergiants ($\beta \approx 1$ and $\alpha \approx 1$) will have more massive and denser CW shell than a binary, where the secondary has a fast tenuous wind such as in main sequence stars ($\beta \approx 1$ and $\alpha \ll 1$).  We also note that $\mcw \propto a$, which suggests that shells in wider binaries might give stronger effect in the collision with SN ejecta.

\subsection{Thickness of the shell}
\label{sec:thick}

\begin{figure*}
    \centering
    \hspace{1.5cm}\includegraphics[width=0.43\textwidth]{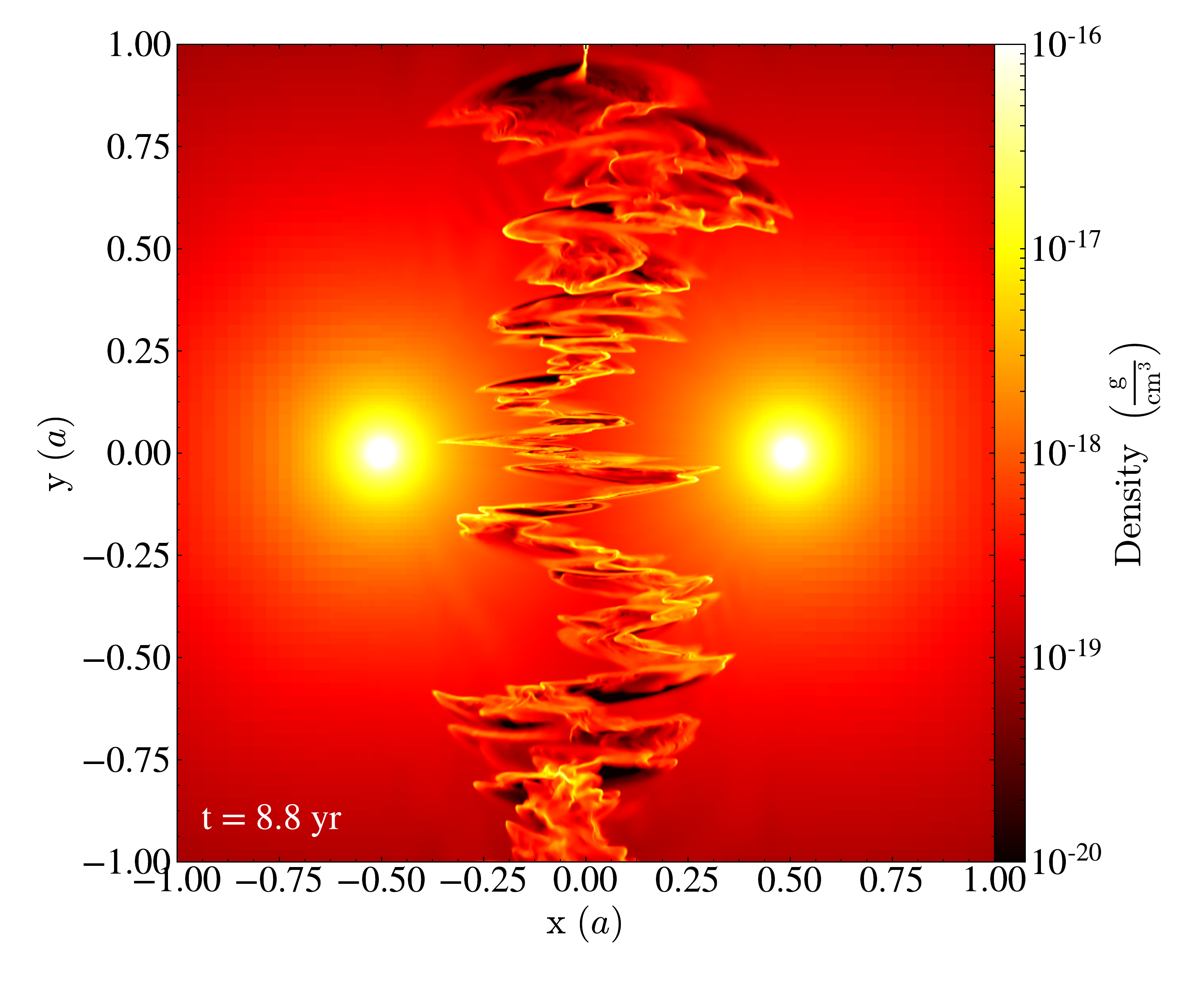}
    \includegraphics[width=0.43\textwidth]{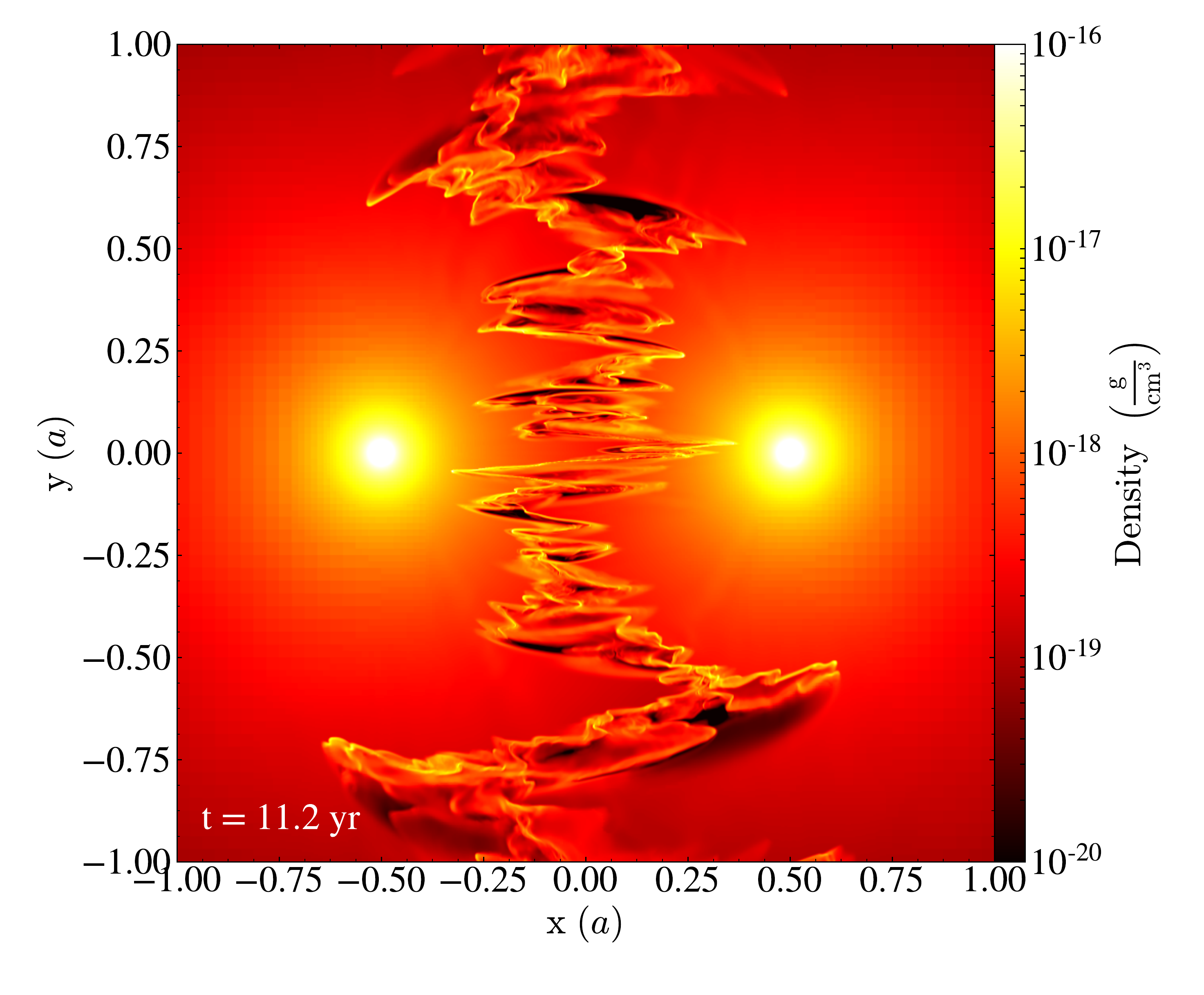}\hfill
    \includegraphics[width=0.8\textwidth]{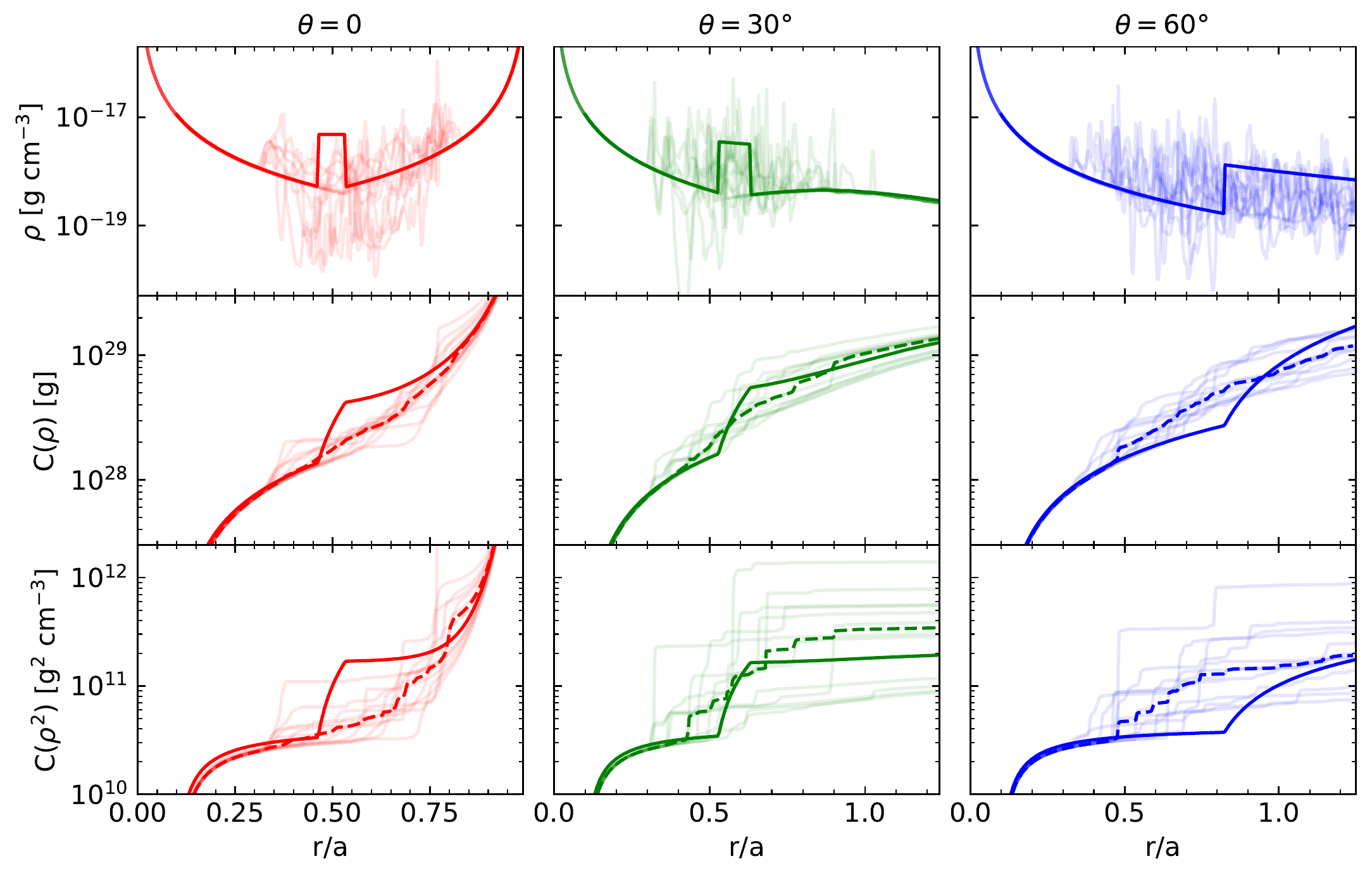}
    \caption{Comparison of hydrodynamical simulations with our analytical model for unstable radiatively-efficient colliding winds. Top row shows density distributions for two identical stars at different times. Bottom part shows profiles of $\rho$ (top), $C(\rho)$ (middle), and $C(\rho^2)$ (bottom) for three different angles $\theta=0$ (left column), $30\degr$ (middle column), and $60\degr$ (right column) measured from the star on the left. Profiles measured from the hydrodynamical simulations at different times are shown with thin lines, their median is shown with dashed line, and our analytical model based on Eq.~(\ref{eq:Delta}) is displayed with a solid line.}
    \label{fig:ntsi}
\end{figure*}

\begin{figure*}
    \centering
    \hspace{1.5cm}\includegraphics[width=0.43\textwidth]{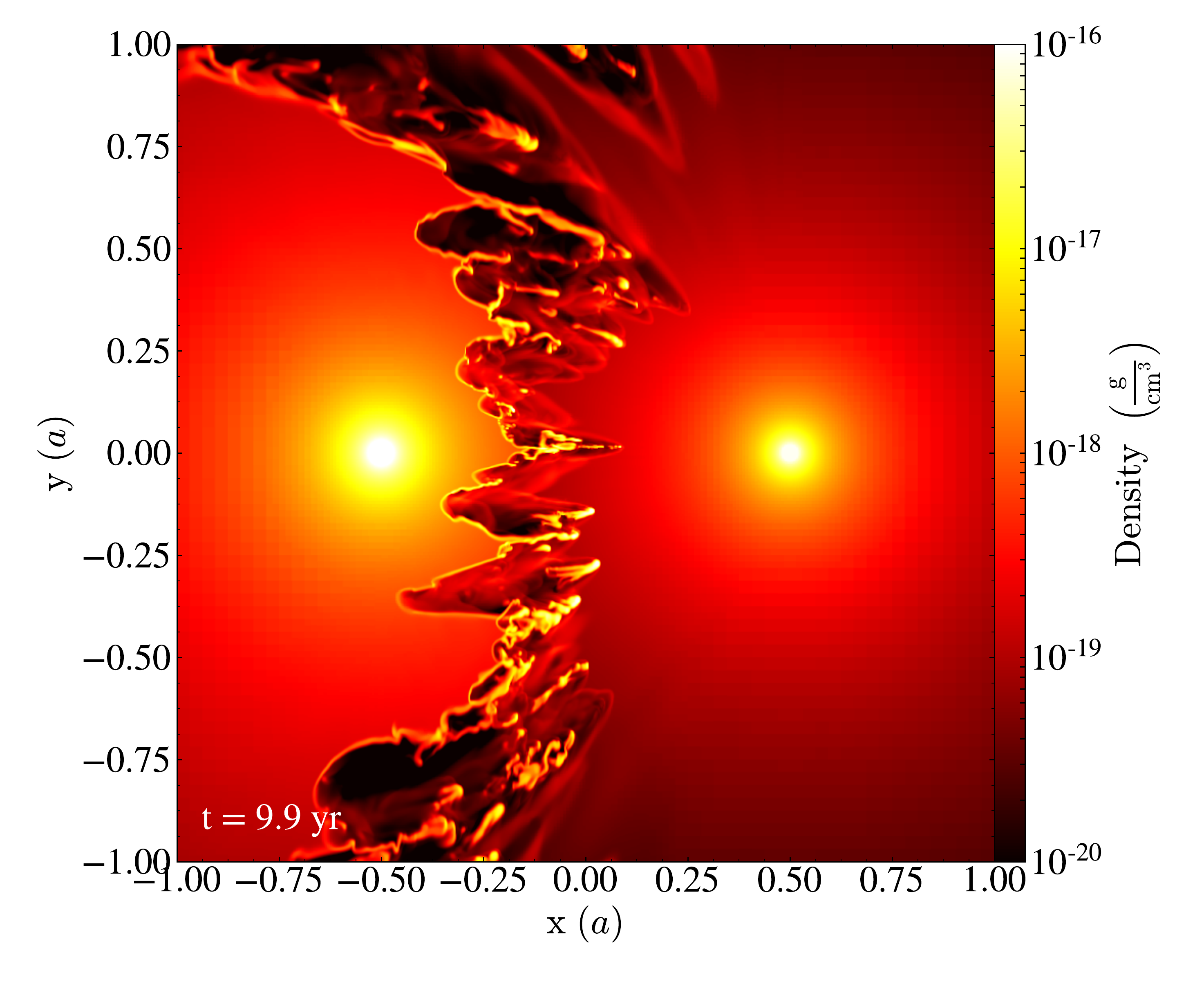}
    \includegraphics[width=0.43\textwidth]{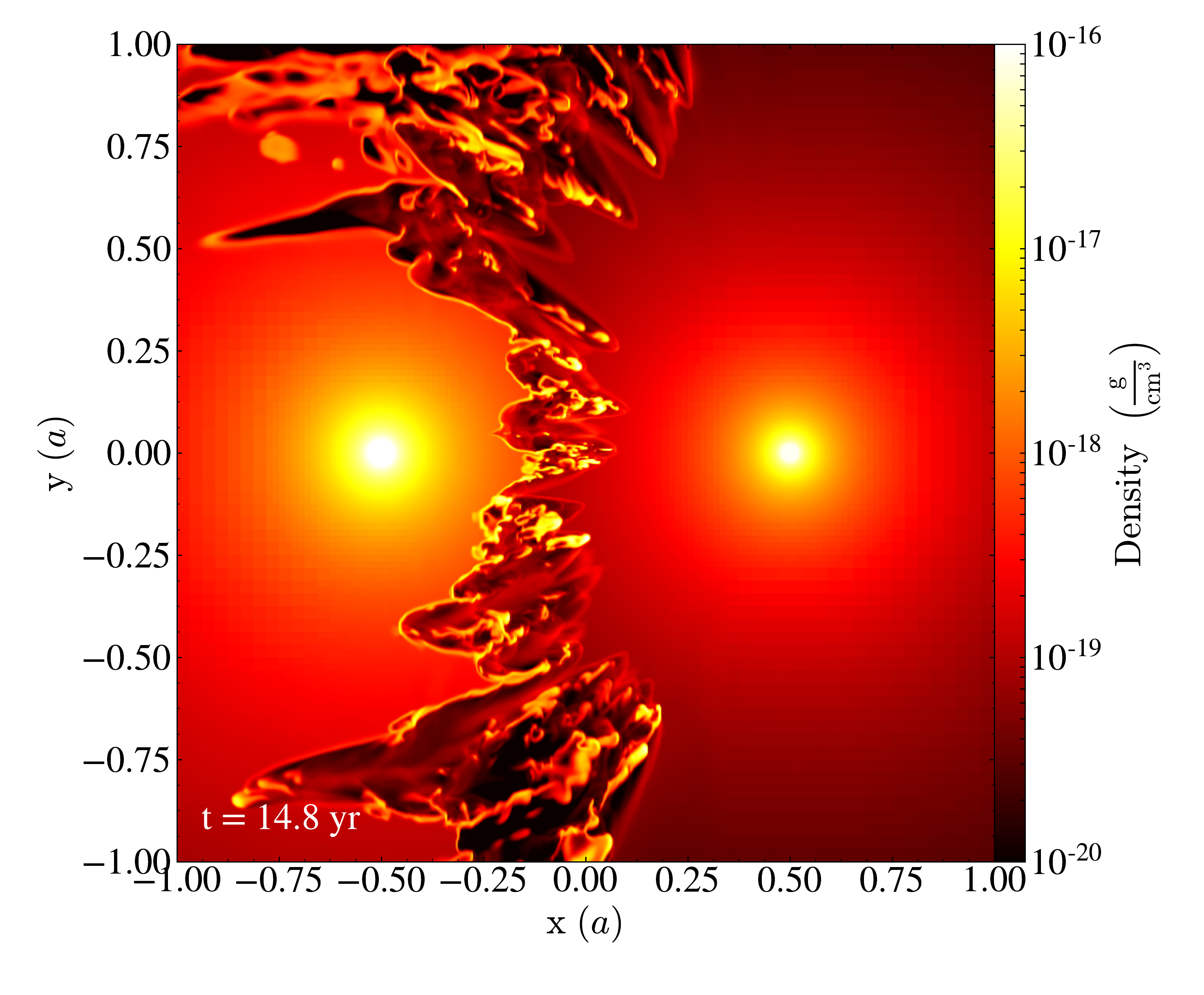}\hfill
    \includegraphics[width=0.8\textwidth]{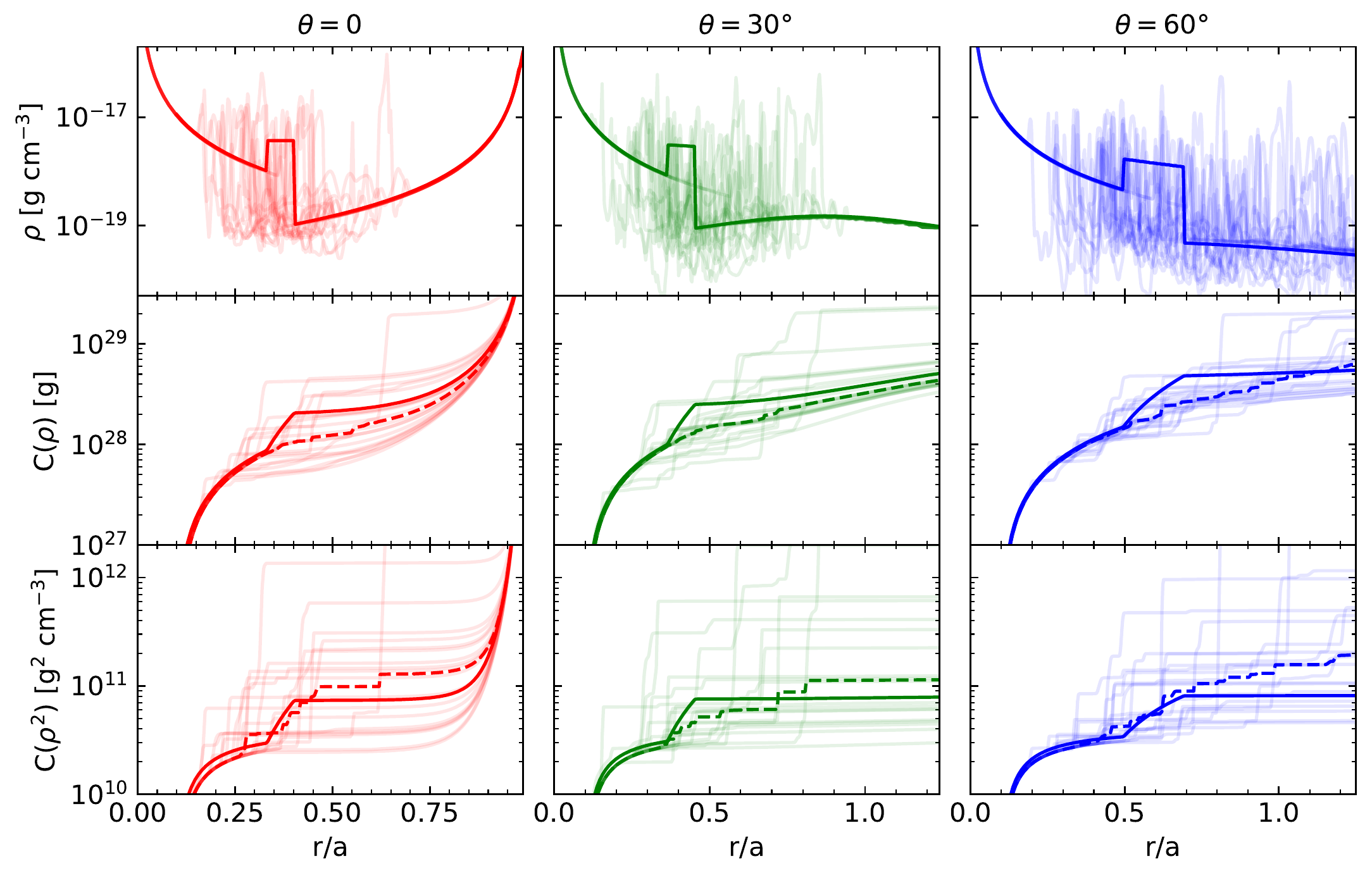}
    \caption{Same as Fig.~\ref{fig:ntsi}, but for unequal stellar winds. Profiles are measured starting from the star on the left.}
    \label{fig:ntsi_unequal}
\end{figure*}

The density inside the CW shell is
\begin{equation}
    \rho_\text{cw} = \frac{\sigma}{2\Delta},
    \label{eq:rhocw}
\end{equation}
where the shell thickness $\Delta$ depends primarily on the efficiency of radiative cooling in the shocked region. 
We constrain the value of $\Delta$ with the aid of three-dimensional hydrodynamical simulations of stellar wind collisions based on the works of \cite{calderon20b,calderon20a}. 
We use the hydrodynamical code \textsc{ramses} \citep{teyssier02} with its AMR module for enhancing the resolution only in certain regions of the domain according to physical criteria. 
First, we simulate adiabatic models, which assumes that radiative cooling is inefficient. 
The domain was set to a cube made out of $64^3$ cells, allowing for up to 4 levels of refinement, which is equivalent to a potential maximum resolution of $1024^3$ cells. 
The refinement strategy was based on density gradients.
We set up the problem by placing two stars fixed in space and each of them emitting a spherically-symmetric wind. 
We place the stars on the plane $z=0$ with a stellar separation of $a=206$\,au. 
We parameterize the winds by their mass-loss rates, terminal velocities, and temperatures. 
We assume that the winds are ejected at their terminal velocities. 
Models with identical stars adopted $\mdot_A = \mdot_B = 10^{-5}\,\myr$, $v_{\infty,A} = v_{\infty,B} =500\rm\ km\ s^{-1}$, and initial wind temperatures of $T=10^4\rm\ K$. 
In addition, we simulated unequal winds where we set $\mdot_B =10^{-4}\,\myr$, i.e. $\beta=0.1$. 
Table~\ref{tab:sims} summarizes the parameters of each model. 
We run the simulations for 10 crossing time-scales $t_{\rm cross}=a/v_\infty$ and we perform the analysis only once the system has reached steady state. 
For more details on the setup, parameters, and a description on the time evolution of the models we refer the reader to \cite{calderon20b}.

\begin{table}
    \centering
    \begin{tabular}{|c|c|c|c|c|}
         \hline
         Type       & $\alpha$ &   $\beta$ & $L/a$ & Max. Res. \\
         \hline
         \hline
         Adiabatic  & 1     &   1       &   10  & $1024^3$\\
         Adiabatic  & 1     &   0.1     &   10  & $1024^3$\\
         Cooling    & 1     &   1       &   2   & $2048^3$\\
         Cooling    & $1/3$ &  $1/3$    &   2   & $1024^3$\\
         \hline
    \end{tabular}
    \caption{Parameters of the numerical models. 
    All simulations used $\dot{M}_{\rm A}=10^{-5}\,\myr$, $v_{\infty,\rm A}=500\rm\ km\ s^{-1}$, and $a=206$\,au
    Column 1: type of model is either adiabatic (without radiative cooling) or cooling (enabled radiative cooling). Column~2: wind terminal velocity ratio $\alpha$. 
    Column~3: wind momentum ratio $\beta$. 
    Column~4: size of the side of the cubic domain. 
    Column~5: maximum potential resolution achieved with AMR.} 
    \label{tab:sims}
\end{table}

We show the resulting density distributions for the adiabatic simulations in the top row of Figure~\ref{fig:adiabatic}, where we see that the density distribution inside the shell is smooth with sharp boundaries.
In the bottom row of Figure~\ref{fig:adiabatic}, we show density profiles along radial rays originating at $A$. By experimenting with various prescriptions, we find that a good match to hydrodynamical simulations can be obtained by setting
\begin{equation}
    \frac{\Delta}{a} = \delta_\text{cw}\sqrt{1 + (3\ell)^{2.2}},
    \label{eq:Delta}
\end{equation}
where $\delta_\text{cw} = 0.07$ for the adiabatic simulations, and $\ell$ is the length of the CW shell arc integrated from $\theta=0$. It is striking how well this prescription matches the numerical simulation of the identical winds (see left column of Fig.~\ref{fig:adiabatic}), including the radial decrease of $\rho_\text{cw}$ inside the shell caused by the ray traversing the shell through different values of $\ell$. The agreement between simulation and model in the case of $\beta=0.1$ is somewhat worse, especially for $\theta=90\degr$, but we still consider this satisfactory since we do not want to overly complicate our analytic model with additional parameters.

Under certain conditions, the shocked shell between the two CW can radiatively cool and become unstable to the thin shell instability \citep{vishniac83,vishniac94}. Numerical simulations  show that the shell collapses to a thin sheet, which buckles into a clumpy corrugated time-dependent structure \citep[e.g.][]{stevens92,pittard09,kee14,steinberg18,calderon20b,calderon20a}. 
In order to analyse systems with such unstable shells we develop a set of simulations with optically-thin radiative cooling enabled.  We also decreased the domain size to a cube with side $2a$ to resolve finer structures. 
Similarly to the adiabatic case, we run one model with identical stars and a second model where one of the winds is three times faster than the other. 
Table~\ref{tab:sims} shows the parameters of the models. 
We show the density distributions at two different simulation times for equal winds in the top panel of Figure~\ref{fig:ntsi} and for the unequal winds in the top panel of Figure~\ref{fig:ntsi_unequal}. We see that the instability fractures the shell and creates regions that can have either higher or lower density than what we would expect from adiabatic simulation (see Fig.~\ref{fig:adiabatic}).

In the bottom part of Figures~\ref{fig:ntsi} and \ref{fig:ntsi_unequal} we see the wide variation in the density profiles at different simulation times shown as thin solid lines. As a result, we can only describe the mean properties of the simulation. We define cumulative distribution function $C(\psi)$ of quantity $\psi$ along a radial cone originating from $A$
\begin{equation}
    C(\psi) = \int^r 4\pi r'^2 \psi dr',
    \label{eq:cdf}
\end{equation}
where the lower integration limit is set close to $A$. We show $C(\rho)$ and $C(\rho^2)$ in the bottom part of Figures~\ref{fig:ntsi} and \ref{fig:ntsi_unequal} to illustrate the mean behavior of quantities that are most relevant for propagation of radiative shocks. Here, we assume that quantities measured along rays in the simulation are representative of values in cones; doing the proper averaging would only lead to less dispersion at larger $r$. We find that the mean behavior of hydrodynamical simulations is relatively well reproduced by assuming effective shell thickness described by Equation~(\ref{eq:Delta}), but with $\delta_\text{cw} = 0.035$. We note that simulation profiles of $\rho(r)$ span much greater range in the CW shell, but only very few realizations achieve higher values than our prescription. This means that were are likely not significantly under-estimating quantities like recombination luminosities.

\begin{figure}
    \centering
    \includegraphics[width=0.48\textwidth]{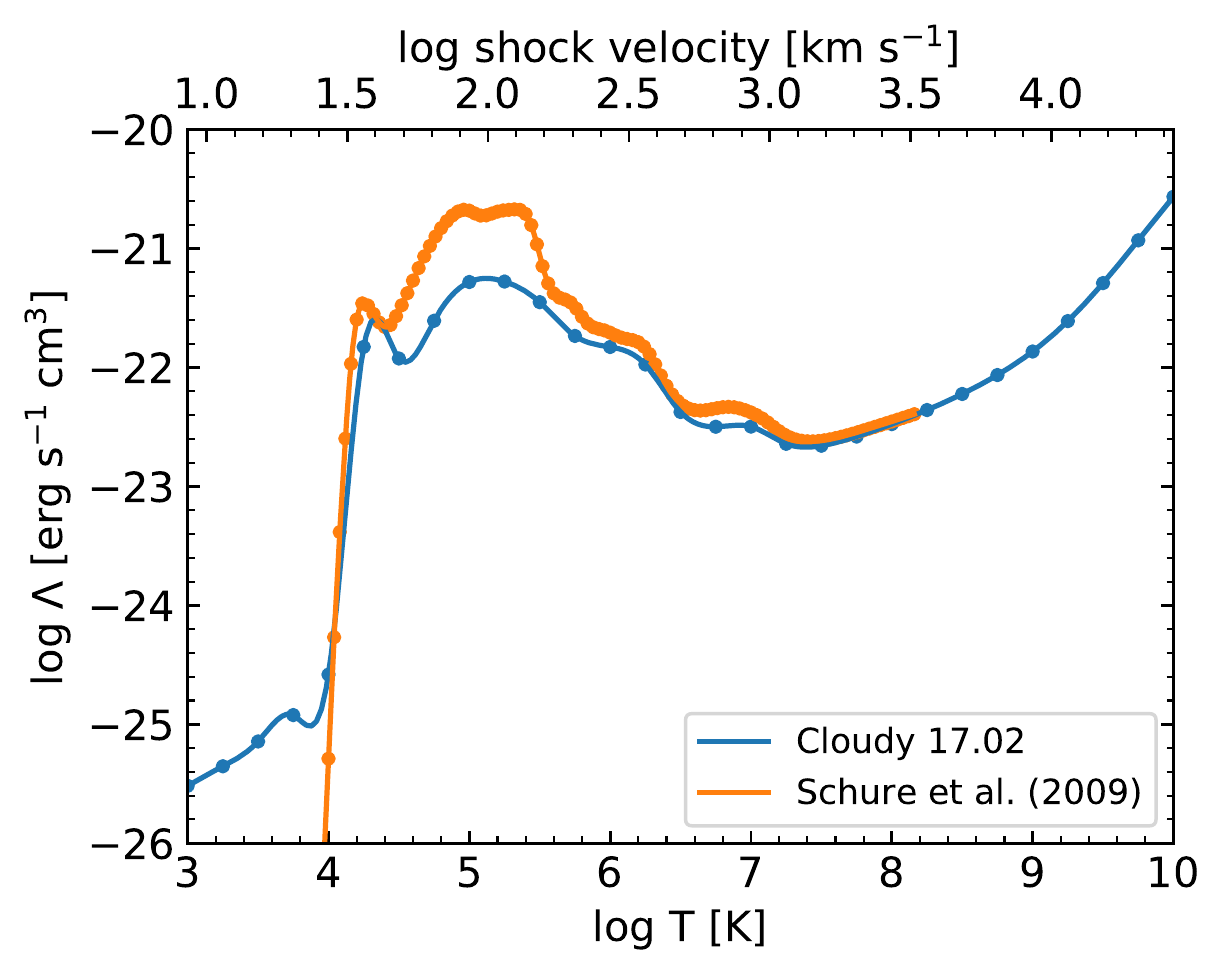}
    \caption{Cooling curve used in this paper. Blue points and line show calculation from Cloudy 17.02 \citep{cloudy17}, which we compare with the cooling curve from  \citet{schure09} shown in orange. The correspondence between shock velocity and temperature is calculated using $T = (3/16) (m_\text{p}/k_\text{B}) v_{\infty}^2$.} 
    \label{fig:cloudy}
\end{figure}

Finally, we need to determine which CW shells are adiabatic and which are radiatively efficient to choose the proper value of $\delta_\text{cw}$. Following \citet{stevens92} and \citet{calderon16}, we define for each stellar wind the cooling parameter $\chi$ as the ratio of the cooling $t_{\rm cool}$ and the dynamical $t_{\rm dyn}$ time-scales, 
\begin{equation}
    \chi = \frac{t_\text{cool}}{t_\text{dyn}} = \left(\frac{3k_\text{B}T}{2 \tilde{\rho} \Lambda(T)/m_\text{p}} \right) \left(\frac{d}{c_S}\right)^{-1},
    \label{eq:chi}
\end{equation}
where $\tilde{\rho} = 4\rho_{A,B}$ is the immediate post-shock density assuming a strong shock, $T = (3/16) (m_\text{p}/k_\text{B}) v_{\infty}^2$ is the immediate post-shock temperature \citep{lamers99}, $m_\text{p} = 1\times 10^{-24}$\,g is the mean particle mass for solar composition, $k_{\rm B}$ is the Boltzmann constant, $c_S^2 = (5/3)(k_\text{B}/m_\text{p}) T$ is the post-shock sound speed, $\Lambda(T)$ is the cooling function, and $d$ is the distance from the star, which is either $d=R_\text{cw,0}$ for $A$ or $d=a-R_\text{cw,0}$ for $B$. We calculate $\Lambda(T)$ in Cloudy 17.02 \citep{cloudy17} by extending the temperature range of the test problem {\tt grid\_h2coronal} to $T=10^{10}$\,K. This cooling curve covers also temperatures $T<10^4$\,K, which is important for properly characterizing collisions of slow winds. We show our cooling curve in Figure~\ref{fig:cloudy}, where we compare it to a cooling curve of \citet{schure09}, which covers smaller range of temperatures. We assume $\delta_\text{cw} = 0.035$ if $\chi_A,\chi_B < 0.1$, and $\delta_\text{cw} = 0.07$ otherwise.  The relatively small difference between the adiabatic and radiatively-efficient winds imply that our results are not very sensitive to assumptions on the wind classification. For many colliding stellar winds, especially those originating from cool extended stars, the assumption of a completely ionized ideal gas that is used to calculate $\chi$ are not valid. Nonetheless, any error in the classification based on $\chi$ will be relatively small, because our estimates of $\delta_\text{cw}$ are similar in both cases.

\subsection{Final density distribution including orbital motion}
\label{sec:density}

To obtain the final CSM density distribution $\rho_\text{CSM}(r,\theta)$, we first calculate the density distribution without taking into account the binary orbital motion, $\rho'_\text{CSM}(r,\theta)$. Calculation of $\rho'_\text{CSM}$ combines density distributions of the CW shell and of the two stellar winds. We approximate the CW shell as a piecewise-linear curve with points that are adaptively distributed to satisfy constraints on their mutual distance and difference in $\theta$. For each point on the CW shell, we evaluate $\sigma$, $\ell$, $\Delta$, and the time $t_\text{cw}$ for the gas to flow inside the shell to the given point from $\theta = 0$. If the distance between an arbitrary point $(r,\theta)$ and the CW shell is smaller than $\Delta$, $\rho'_\text{CSM}$ is set to $\rho_\text{cw}$ corresponding to the nearest point on the CW shell. In the remaining cases, $\rho'_\text{CSM}$ is set to either $\rho_A$ or $\rho_B$, depending on which side of the shell the point is located.

The final ingredient of the model is to take into account the binary orbital motion. This effect removes the axial symmetry of the problem and deforms the CW shell. We do not try to model the change in the shape of the shell, but instead assume that the shell geometry is unchanged inside a critical radius $r_\text{mix}$ and that outside of $r_\text{mix}$ the CSM density smoothly transitions to a blend of the two winds assuming momentum conservation,
\begin{equation}
    \rho_\infty = \frac{\mdot_A + \mdot_B}{4\pi r^2 v_{\infty}},
\end{equation}
where the asymptotic velocity $v_\infty$ is
\begin{equation}
v_\infty = \frac{\mdot_A v_{\infty,A} + \mdot_B v_{\infty,B}}{\mdot_A + \mdot_B}.
\end{equation}
We determine $r_\text{mix}$ by equating $t_\text{cw}$ and the binary orbital period $P$. The inner and outer density and velocity distributions are combined to give the final density distribution
\begin{equation}
    \rho_\text{CSM} = w \rho'_\text{CSM} + (1-w) \rho_\infty,
\end{equation}
where the interpolation coefficient is
\begin{equation}
    w = \left[1+\exp\left(\frac{r-r_\text{mix}}{0.25a}\right)\right]^{-1}.
\end{equation}
A similar algorithm is applied to obtain the CSM radial velocity as seen from star $A$,  $v_\text{CSM}$. Our implementation of the orbital motion in our semi-analytic model  remains to be verified by a hydrodynamical simulation.

\section{Model of light curves}
\label{sec:lc}

Here, we present our model for calculating the shocked shell dynamics (Section~\ref{sec:shock_dynamics}), radiated shock power (Section~\ref{sec:shock_radiation}), and the recombination light curves (Section~\ref{sec:recombination}). We model the shock as a thin shell, which evolves according to mass and momentum conservation laws in the presence of spherical SN ejecta and aspherical CSM. As a result, the thin shell properties change as function of direction. We model this by considering the dynamics along radial cones with different $\theta$. With this exception, our model closely follows \citet{metzger14}. 

\subsection{Shocked shell dynamics}
\label{sec:shock_dynamics}
The evolution of shocked shell mass per solid angle $\msh(\theta)$, shell velocity $\vsh(\theta)$, and shell radius $\rsh(\theta)$ is described by 
\begin{eqnarray}
\frac{\text{d} \msh}{\text{d} t} &=& \rsh^2\left[ \rhosn (v_\text{SN} - \vsh) +  \rhocsm(\vsh-v_\text{CSM})\right],\\
\frac{\text{d}\vsh}{\text{d} t} &=&  \frac{\rsh^2}{\msh} \left[\rhosn (v_\text{SN} - \vsh)^2 - \rhocsm (\vsh-v_\text{CSM})^2\right],\\
\frac{\text{d}\rsh}{\text{d} t} &=& \vsh,
\end{eqnarray}
where $\rho_\text{CSM}$ and $v_\text{CSM}$ are defined in Section~\ref{sec:csm}, and $\rhosn$ and $v_\text{SN}$ are density and velocity of SN ejecta given by the prescription of \citet{chevalier89} 
\begin{equation}
\rhosn = 
\begin{dcases}
\rho_\text{tr} \left(\frac{r_\text{tr}}{r}\right)^{\delta_\text{SN}}, & r < r_\text{tr}\\
\rho_\text{tr} \left(\frac{r_\text{tr}}{r}\right)^{n_\text{SN}}, & r_\text{tr} \le r \le r_\text{SN}\\
0, & r > r_\text{SN},
\end{dcases}
\end{equation}
where $r_\text{tr} = v_\text{tr} t$ is the transition radius between the power laws with slopes $\delta_\text{SN} = 0.5$ and $n_\text{SN} = 12$, and $r_\text{SN} = v_\text{max} t$ is outermost SN radius with $v_\text{max} = 10 v_\text{tr}$. Density and velocity at $r_\text{tr}$ are given by
\begin{eqnarray}
    \rho_\text{tr} &=& \frac{(3-\delta_\text{SN})(n_\text{SN} - 3)}{4\pi (n_\text{SN}-\delta_\text{SN})} \frac{M_\text{SN}}{r_\text{tr}^3},\\
    v_\text{tr} &=& \left[\frac{2(5-\delta_\text{SN})(n_\text{SN}-5)}{(3-\delta_\text{SN})(n_\text{SN}-3)}\frac{E_\text{SN}}{M_\text{SN}}\right]^{1/2},
\end{eqnarray}
where we take SN explosion energy $E_\text{SN} = 10^{51}$\,erg and total ejecta mass $M_\text{SN} = 10\,\msun$, which gives $v_\text{tr} \approx 3700\,\kms$. \citet{kurfurst20} found that assuming realistic density profile instead of broken power law leads to only small changes in the hydrodynamical evolution.

\subsection{Radiated shock power}
\label{sec:shock_radiation}

We now calculate the fraction of shock power released in electromagnetic radiation. The shell converts mechanical to thermal energy at reverse and 
forward shocks with rates
\begin{equation}
\dot{Q}_\text{r} \approx \frac{3}{8} P_\text{r}\rsh^2 (\rsh/t - \vsh),\quad \dot{Q}_\text{f} \approx \frac{3}{8} P_\text{f}\rsh^2 (\vsh - v_\text{CSM}),
\end{equation}
where the post-shock pressures are given by 
\begin{equation}
P_\text{r} \approx \rhosn(\rsh/t - \vsh)^2,\quad P_\text{f} \approx \rhocsm. (\vsh-v_\text{CSM})^2
\end{equation}
In this expression, we neglect the difference between adiabatic and radiative shocks and assume the pressures are approximately average of these two cases. This difference from our assumed value should be about $30$~per~cent \citep{metzger14}. We also assume complete ionization and an ideal gas with adiabatic index of $5/3$. This is a good approximation, because the SN shock will travel with velocities of a few thousand km\,s$^{-1}$.

The radiative efficiencies of the two shocks are given by
\begin{equation}
\eta_\text{r} = \frac{P_\text{r}}{n_\text{r}^2 \Lambda(T_\text{r})} \frac{\vsh}{\rsh},\quad \eta_\text{f} = \frac{P_\text{f}}{n_\text{f}^2 \Lambda(T_\text{f})} \frac{\vsh}{\rsh},
\end{equation}
where the post-shock temperatures are given by $T = P/(k_\text{B} n)$, and the post-shock densities are calculated assuming strong shocks in a medium with adiabatic index of $5/3$, 
\begin{equation}
n_\text{r} = 4\frac{\rhosn}{m_\text{p}},\quad n_\text{f} = 4\frac{\rhocsm}{m_\text{p}}.
\end{equation}
The final radiated shock powers are given by interpolation between adiabatic and radiative regimes as
\begin{equation}
L_\text{r} = \frac{\dot{Q}_\text{r}}{1 + (5/2)\eta_\text{r}},\quad L_\text{f} = \frac{\dot{Q}_\text{f}}{1 + (5/2)\eta_\text{f}},
\label{eq:Lf}
\end{equation}
where we use the cooling function $\Lambda(T)$ shown in Figure~\ref{fig:cloudy}. The final radiative power emitted by the system is obtained by integrating over $\theta$ as 
\begin{equation}
\Lshock = 2\pi\int_0^\pi (L_\text{r} + L_\text{f})\sin\theta    
\,
\text{d}\theta.
\end{equation}
We evaluate the integral at a non-uniform grid of $\theta$ designed with constraints on the coverage of the CW shell. We take into account directions with $\theta > \theta_\infty$, where the SN ejecta does not encounter the CW shell and where the interaction with effectively single-star wind contributes to $\Lshock$. Typically, we use more than $100$ values of $\theta$.

For typical SN ejecta velocities, the luminosity $\Lshock$ is emitted in UV a X-rays. Some or all of the radiation can be reprocessed to optical wavelength if the shock interaction region is surrounded by sufficiently dense medium, which can be either the dense thin shell itself, the yet unshocked CSM, or even the SN ejecta that can wrap around and envelope certain non-spherical shock interaction regions \citep[e.g.][]{smith15,suzuki19,kurfurst19,kurfurst20}. In such cases, the radiation needs to diffuse through the reprocessing layer, which typically makes the emission last longer and reduces the peak luminosities \citep[e.g.][]{arnett82,chatzopoulos12}. Our subsequent presentation ignores this complication, because we are interested in assessing the maximum possible effect of shock interaction with CW shells. Our results should thus be viewed as optimistic.

There are additional complications that our model does not take into account. For example, the individual directions  are treated independently, yet we know that hydrodynamic instabilities effectively couple nearby angles \citep{kurfurst20}. Dust formation can occur in the radiatively-cooled shell, which might block some of the radiation. Differences from solar composition will affect the cooling efficiency. This would be particularly important for colliding winds in Wolf-Rayet binaries.

\subsection{Recombination light curves}
\label{sec:recombination}
 
The goal of our treatment is to assess the dependence of flash ionization signature on binary and wind parameters, and to see whether this dependence differs from radiated shock luminosity. Following \citet{kochanek19}, we calculate the relative recombination rate per solid angle as
\begin{equation}
    \frac{\gamma(t)}{\gamma_\infty} = \int_{\rsh}^{R_\text{bo}} r^2 \rho_\text{CSM}^2
\,
\text{d}r, 
\end{equation}
where $R_\text{bo} = R_A + c t$ is the distance from star $A$ that was reached by shock breakout photons. The normalizing factor $\gamma_\infty$ is
\begin{equation}
    \gamma_\infty = \frac{\mdot_A^2}{(4\pi v_{\infty,A})^2 R_A}.
\end{equation}
The total recombination rate is
\begin{equation}
\frac{\Gamma(t)}{\Gamma_\infty} = \frac{1}{2}\int_0^\pi \frac{\gamma(t)}{\gamma_\infty} \sin\theta
\,
\text{d}\theta,
\end{equation}
where the normalizing factor is $\Gamma_\infty = 4\pi\gamma_\infty$. 

Our treatment improves over \citet{kochanek19} by using a more realistic model of the CSM and by self-consistently calculating $\rsh$. Unlike \citet{kochanek19}, we do not aim to calculate light curves that could be directly compared to observations. The most significant omission of our model are light travel effects, which are of the order of $0.5 (a/$100\,au)\,days. In principle, these effects could be added to our model, but they were already characterized by \citet{kochanek19}.

\section{Results}
\label{sec:results}

Here, we present results of our model combining radiative evolution of thin shocked shells with CSM distribution appropriate for CW binaries. In Section~\ref{sec:basic}, we show evolution of basic quantities like $\msh$, $\vsh$, $\rsh$, and $\Lshock$ as a function of angle $\theta$. In Section~\ref{sec:simple}, we explore the angle-integrated evolution of $\Lshock$ for several model cases of stellar binaries and investigate the dependence on their parameters. In Section~\ref{sec:flash}, we calculate flash ionization signatures. In Section~\ref{sec:pop}, we quantify the relative rates based on known binary statistics.

\subsection{Shocked shell properties as a function of $\theta$}
\label{sec:basic}

\begin{figure*}
    \centering
    \includegraphics[width=0.7\textwidth]{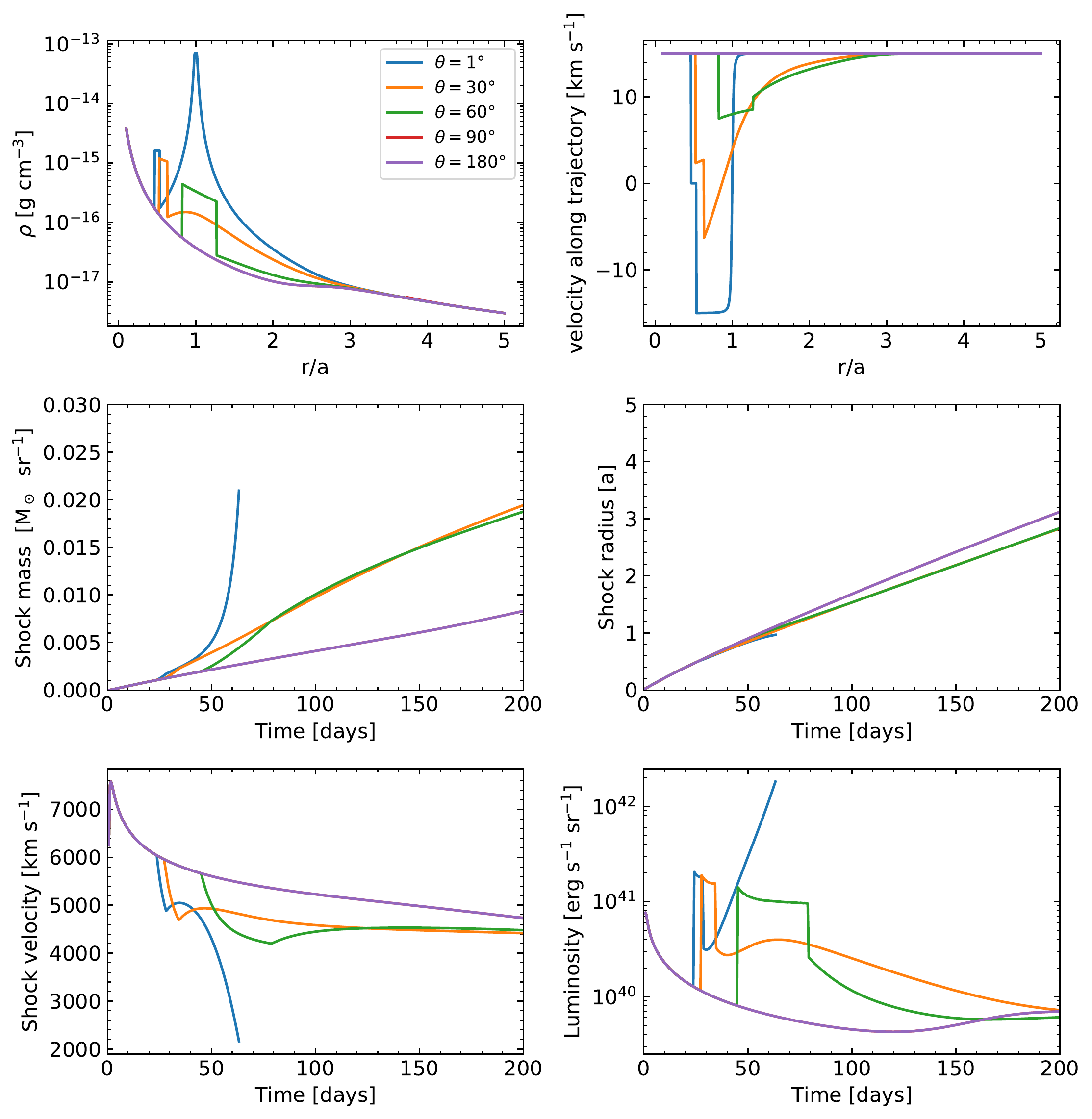}
    \caption{Quantities relevant for our model of shock interaction between SN ejecta and a CW shell. The binary parameters were $M_A=M_B=20\,\msun$, $R_A=R_B=1000\,\rsun$, $\mdot_A=\mdot_B=10^{-4}\,\myr$, $v_{\infty,A}=v_{\infty,B}=15\,\kms$, and $a=200$\,au. We show radial profiles of density (top left) and  CSM radial velocity (top right), and time evolution of mass in the shocked shell (middle left), shock radius (middle right), shock velocity (bottom left), and the shock luminosity (bottom right). All quantities are evaluated along $\theta = 1\degr$ (blue), $30\degr$ (orange), $60\degr$ (green), $90\degr$ (red), and $180\degr$ (purple).}
    \label{fig:basic}
\end{figure*}

In Figure~\ref{fig:basic}, we show results of our model for five directions $\theta$ for a binary composed of two identical red supergiants. In this case, the CW shell is a planar sheet. The density profiles in the top left panel show three features. First, $\rho \propto r^{-2}$  dependence of unperturbed stellar winds leads to two peaks: one at $r\rightarrow R_A$ and the second at $r \approx a$ for small $\theta$. Second, the CW shell overdensities occur in directions with $\theta \le 60\degr$. Third, the density profiles transition to the outer mixed medium at $r\gtrsim 3a$. The CSM radial velocity profiles in the top right panel show three distinct regimes: constant $v_\text{CSM}$ in the domain of star $A$, somewhat lower but still positive $v_\text{CSM}$ in the CW shell, which slightly increases with $r$ due to density variations along the inclined ray, and a more variable velocity in the domain of star $B$.

The shocked shell radii $\rsh$ in the middle right panel evolve similarly for all angles, but the accumulated mass per solid angle $\msh$ in the middle left evolves quite differently. Steepest increase is seen for the smallest $\theta$ (for numerical reasons we use $\theta=1\degr$), because this trajectory encounters high-density regions near star $B$; the integration is terminated when the shell reaches the surface of $B$. Trajectory with $\theta=180\degr$ does not encounter the CW shell and $\msh$ remains small. Correspondingly, $\vsh$ in the bottom left panel shows drops when the shocked shell encounters the CW shell. At late time, all trajectories converge to $\vsh$ slightly above $v_\text{tr}$. Shock luminosity is highest when the shocked shell is sweeping up the CW shell, but the peak occurs at different times as a function of $\theta$. Typically, $L_\text{f}$ has a major contribution during the peak while $L_\text{r}$ dominates after exiting the shell. Trajectory with $\theta=1\degr$ has $\Lshock$ steeply increasing close to $B$, but this is inconsequential for the angle-integrated luminosity due to the small solid angle subtended by $B$. 

\subsection{Evolution of shock power}
\label{sec:simple}

\begin{figure*}
    \centering
    \includegraphics[width=0.7\textwidth]{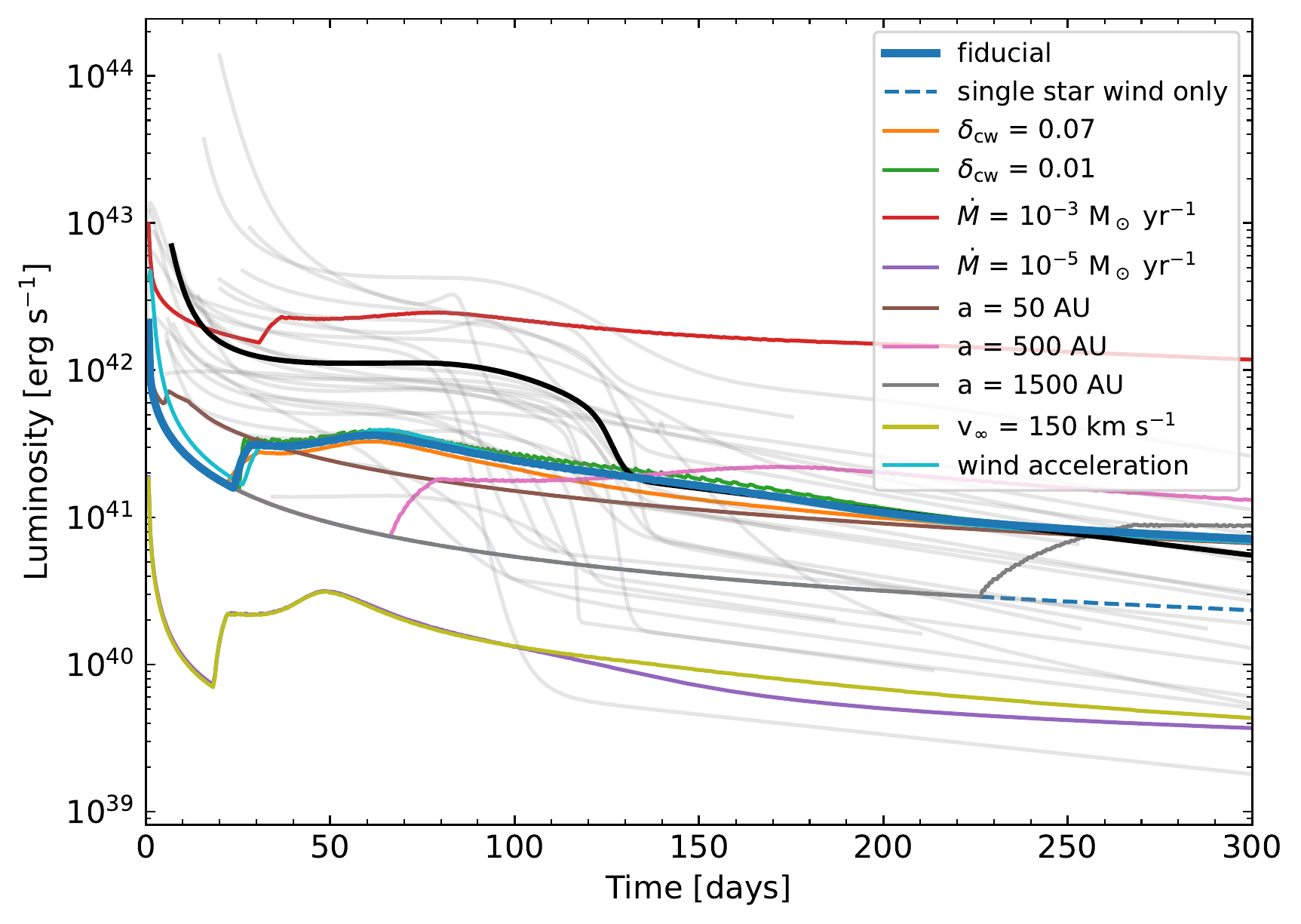}
    \caption{Evolution of shock luminosity when changing one parameter at a time. The fiducial run (thick blue line) is for the same twin red supergiant binary as in Fig.~\ref{fig:basic}. The remaining lines show runs where we change one parameter in the same way for both stars: width of the CW shell $\delta_\text{cw}$ (orange and green), wind mass loss rate $\mdot$ (red and purple), binary separation $a$ (brown, pink, and grey), and wind velocity $v_\infty$ (light green). Blue dashed line shows shock luminosity for the case of SN explosion in a single-star wind. Furthermore, the light blue solid line shows the single-star case but taking the wind acceleration into account. Thin gray lines in the background show bolometric light curves of normal Type II-P supernovae from \citet{pejcha15}. SN1999em is emphasized with a black line.  }
    \label{fig:single_equal}
\end{figure*}

We show the angle-integrated shock light curves in Figure~\ref{fig:single_equal}. The fiducial calculation corresponds to a twin red supergiant binary and shows increase of $\Lshock$ about 25 days after explosion. At peak, the $\Lshock$ is approximately by a factor of $5$ higher than for an otherwise equivalent single-star wind. Higher $\Lshock$ is maintained for hundreds of days as the SN ejecta collide with progressively more distant parts of the shell and as the SN ejecta re-accelerate the shocked shell. The peak luminosity is most influenced by the density in the CW shell, which is proportional to the wind $\mdot$. As expected, CW shells from winds of lower $\mdot$ are less radiatively efficient, which further suppresses $\Lshock$. A similar effect comes from changing the wind velocity $v_\infty$. For example, increasing wind velocity by a factor of 10 is nearly identical to lowering $\mdot$ by the same factor. There are, however, small differences, because the CSM velocity enters in the dynamical equations. Binary separation sets the time when $\Lshock$ starts to increase and also influences the total mass in the CW shell, $M_\text{cw} \propto a$.  We see that varying $a$ by a factor of $4$ changes the peak $\Lshock$, but the effect is relatively small. Our model predicts that the CW shell in the fiducial case is radiatively unstable with shell width $\delta_\text{cw} = 0.035$. We see that changing $\delta_\text{cw}$ by hand to either the adiabatic case or to a much thinner shell has relatively small effect on the peak and time evolution of $\Lshock$.

\begin{figure*}
    \centering
    \includegraphics[width=0.7\textwidth]{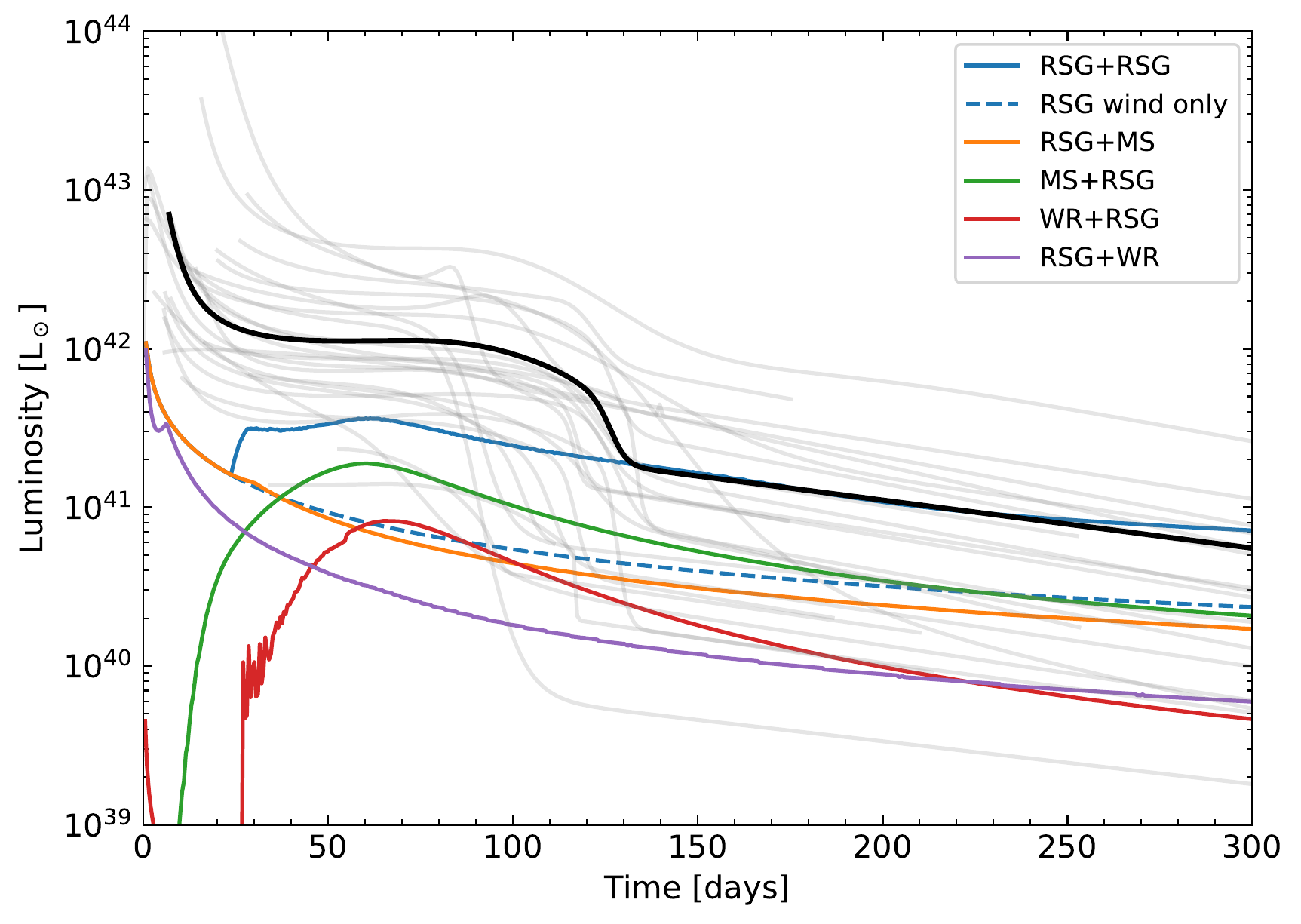}
    \caption{Evolution of shock luminosity for non-equal binaries with $a=200$\,au.  We consider three types of stars: red supergiants (RSG, $\mdot=10^{-4}\,\myr$, $v_\infty = 15\,\kms$, $R=1000\,\rsun$), hot main sequence stars (MS, $\mdot = 10^{-6}\,\myr$, $v_\infty =1500\,\kms$, $R=10\,\rsun$), and Wolf-Rayet stars (WR, $\mdot=10^{-4}\,\myr$, $v_\infty=1500\,\kms$, $R=10\,\rsun$). We show results for different combinations of these stars, as explained in the legend. Blue dashed line shows a single-star wind of a RSG as a reference. Thin gray lines in the background show bolometric light curves of normal Type II-P supernovae from \citet{pejcha15}. SN1999em is emphasized with a black line.}
    \label{fig:single_different}
\end{figure*}

In Figure~\ref{fig:single_different}, we show results for binaries composed of different types of stars, specifically, red supergiants (RSG), hot main sequence stars (MS), and Wolf-Rayet stars (WR). These stars differ by their $\mdot$ and $v_\infty$ and sample different values of $\alpha$ and $\beta$. We see that highest $\Lshock$ is obtained for RSG+RSG binaries, which have slow dense winds. Combining RSG with a MS star with equal wind momentum ($\beta=1$), but different wind velocity ($\alpha=10^{-2}$) gives little change with respect to a single-star wind. For reasons explained in the discussion of Figure~\ref{fig:canto} in Section~\ref{sec:thin_shell}, lowering $\alpha$ gives smaller $M_\text{cw}$ and therefore lower effect on $\Lshock$. Conversely, swapping the labels of the stars in the binary (MS+RSG) leads to $\Lshock$, which is higher at peak than for RSG+MS but still lower than for RSG+RSG. Furthermore, MS+RSG shows a much more clearly defined peak of $\Lshock$. The reason for this behavior is that for MS+RSG the shocked shell first propagates through the tenuous MS wind and encounters CW shell with lower $\msh$. Finally, MS+RSG case might require more complicated binary- or triple-star evolution to have the MS star explode before the RSG.

The situation is similar when combining RSG with WR. WR winds have both fast $v_\infty$ and high $\mdot$, which changes $\beta$. WR+RSG system has $\beta = 10^2$ and $\alpha=10^{-2}$, which gives very small $M_\text{cw}$ and hence low $\Lshock$. For the RSG+WR system, $\beta = 10^{-2}$ and the CW shell is located much closer to RSG. As a result, the peak of $\Lshock$ moves to earlier times and $\Lshock$ is also relatively small. Furthermore, having RSG explode before WR would again require more complicated binary- or triple-star evolution. Based on these results, it is clear that MS+WR, MS+MS, or WR+WR binaries will have even smaller $M_\text{cw}$ and therefore smaller $\Lshock$. We do not show results for these systems here.

In Figures~\ref{fig:single_equal} and \ref{fig:single_different}, we show also bolometric light curves of normal Type II-P SNe from \citet{pejcha15}. Although Type II-P SN radiate primarily in the optical in the first few hundred days and $\Lshock$ comes out primarily in UV or X-rays unless externally reprocessed, it is still useful to compare these light curves to inform the binary parameters that might lead to the strongest observational signature. We find the strongest signature for twin RSG binaries, where the CW shells will influence typical SN plateaus for wind mass loss rates significantly higher than $10^{-4}\,\myr$. Weak SN explosions still require $\mdot \gtrsim 10^{-5}\,\myr$ to have an observable effect on the plateau luminosity. The prospects are better after the plateau ends, because the shock luminosity decreases somewhat slower than the radioactively powered SN light curve. Nonetheless, the requirements on $\mdot$ are still similar. For RSG in a binary with a MS or WR star, the physics of CW shells requires RSG mass-loss rates about a factor of $10$ higher to have similar $\Lshock$ as in the RSG+RSG case. 

Based on our results, we estimate that only in binaries with $a\gtrsim 50$\,au will the shock luminosity rise sufficiently late so that it is not overwhelmed by the bright cooling emission of the SN. Binaries with larger $a$ reach higher shock powers because of the CW shell mass increases with $a$. Furthermore, because the orbital velocity decreases with $a$, the CW shell stays coherent over larger physical scales. At the same time, the density in the shell decreases with $a$, which lowers the radiative efficiency of the shock. The upper limit on $a$ is thus set by the time interval over which are SNe typically observed and by the radiative cooling efficiency. Figure~\ref{fig:single_equal} suggests that for $a>1500$\,au the shock power increases later than $250$\,days after the explosion, when observations are only rarely taken. Since the SN is transparent in the optical at such late times, the shock power would likely manifest in emission lines rather than the continuum.

\subsection{Flash ionization}
\label{sec:flash}

\begin{figure}
    \centering
    \includegraphics[width=0.48\textwidth]{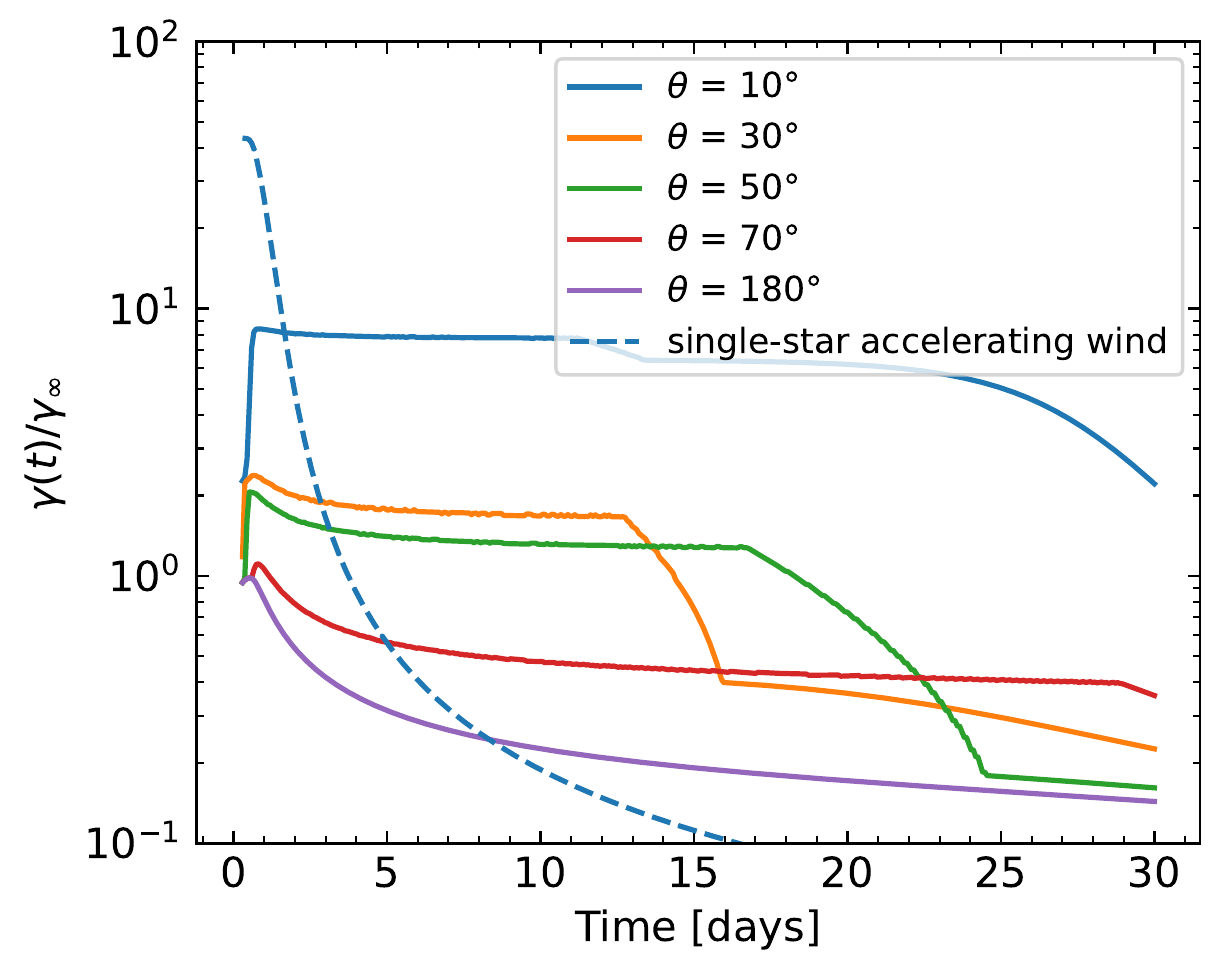}
    \caption{Relative recombination rates per solid angle as a function of direction for a twin binary with $\mdot_A=\mdot_B=10^{-4}\,\myr$, $v_{\infty,A}=v_{\infty,B}=15\,\kms$, and $a=100$\,au. Line color indicates angle $\theta$ explained in the legend. Solid lines correspond to our fiducial model while the blue dashed line is for a modified single-star model that takes into account wind acceleration close to the star with radial exponent of $3.5$ and initial velocity set to $10$~per~cent of the final value. Time delay is not included.}
    \label{fig:gamma_angle}
\end{figure}

In Figure~\ref{fig:gamma_angle}, we show the relative recombination rates $\gamma/\gamma_\infty$ as a function of position angle. As explained by \citet{kochanek19}, there is a peak due to competing effects of increasing $R_\text{bo}$, which increases the outer boundary of the recombination volume, and increasing $\rsh$, which ingests the high-density regions close to the star. For CW shells, this leads to an increase in $\gamma$, which starts when $R_\text{bo} \approx R_{\text{cw},0}$ and which saturates when a substantial fraction of the CW shell is inside $R_\text{bo}$. The recombination rates begin to drop when the shocked shell starts to interact with the CW shell, $\rsh \approx R_{\text{cw},0}$. We see that highest $\gamma$ is reached for low $\theta$, where the CW shell is densest. However, this effect is counteracted by the small solid angle at small $\theta$ and we took special care in the angle integration to take this into account. This effect would be suppressed for companions with more tenuous winds. The profile for  $\theta=180\degr$ is very similar to a single-star wind, because there is no CW shell in this direction.

\begin{figure}
    \centering
    \includegraphics[width=0.48\textwidth]{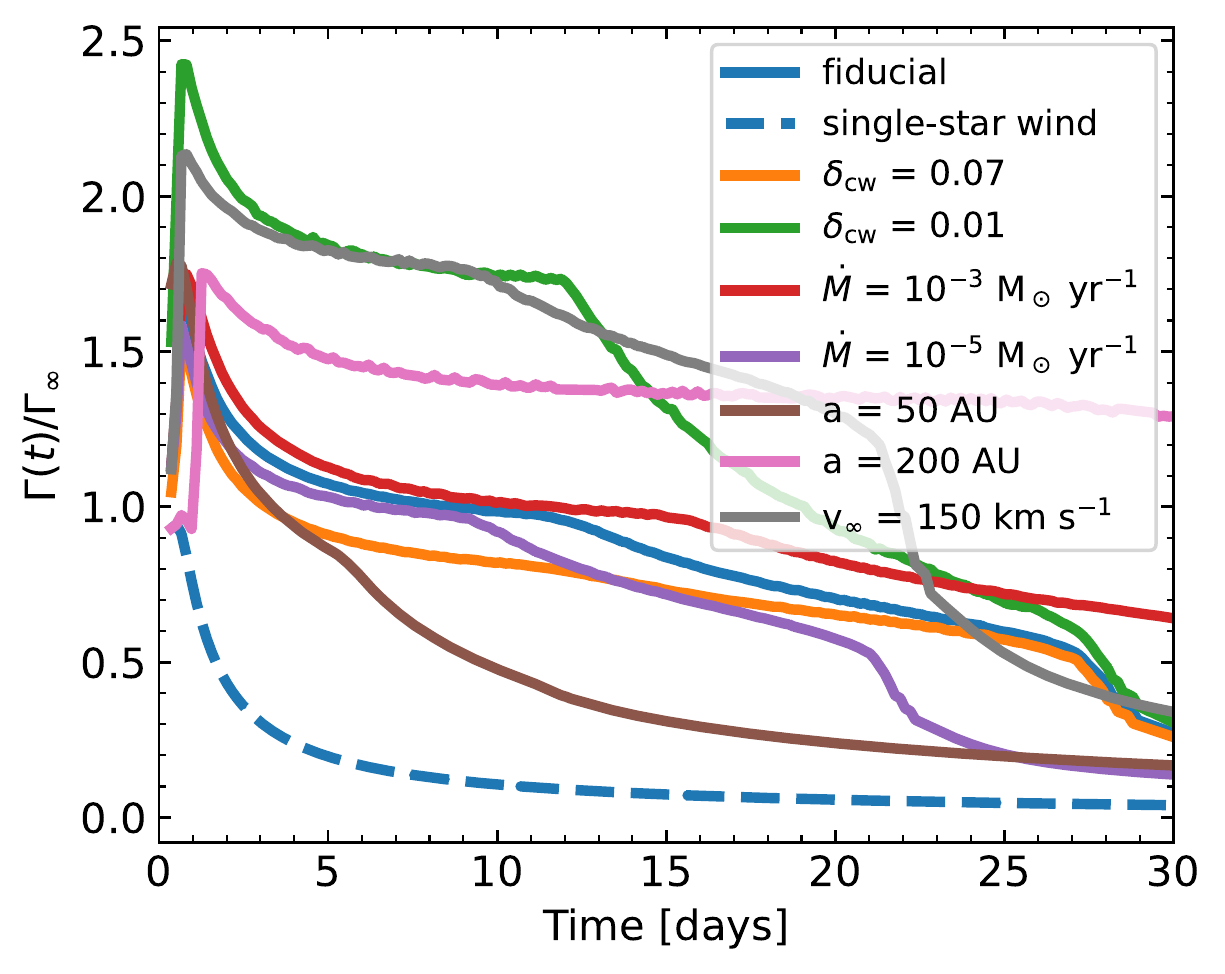}
    \includegraphics[width=0.48\textwidth]{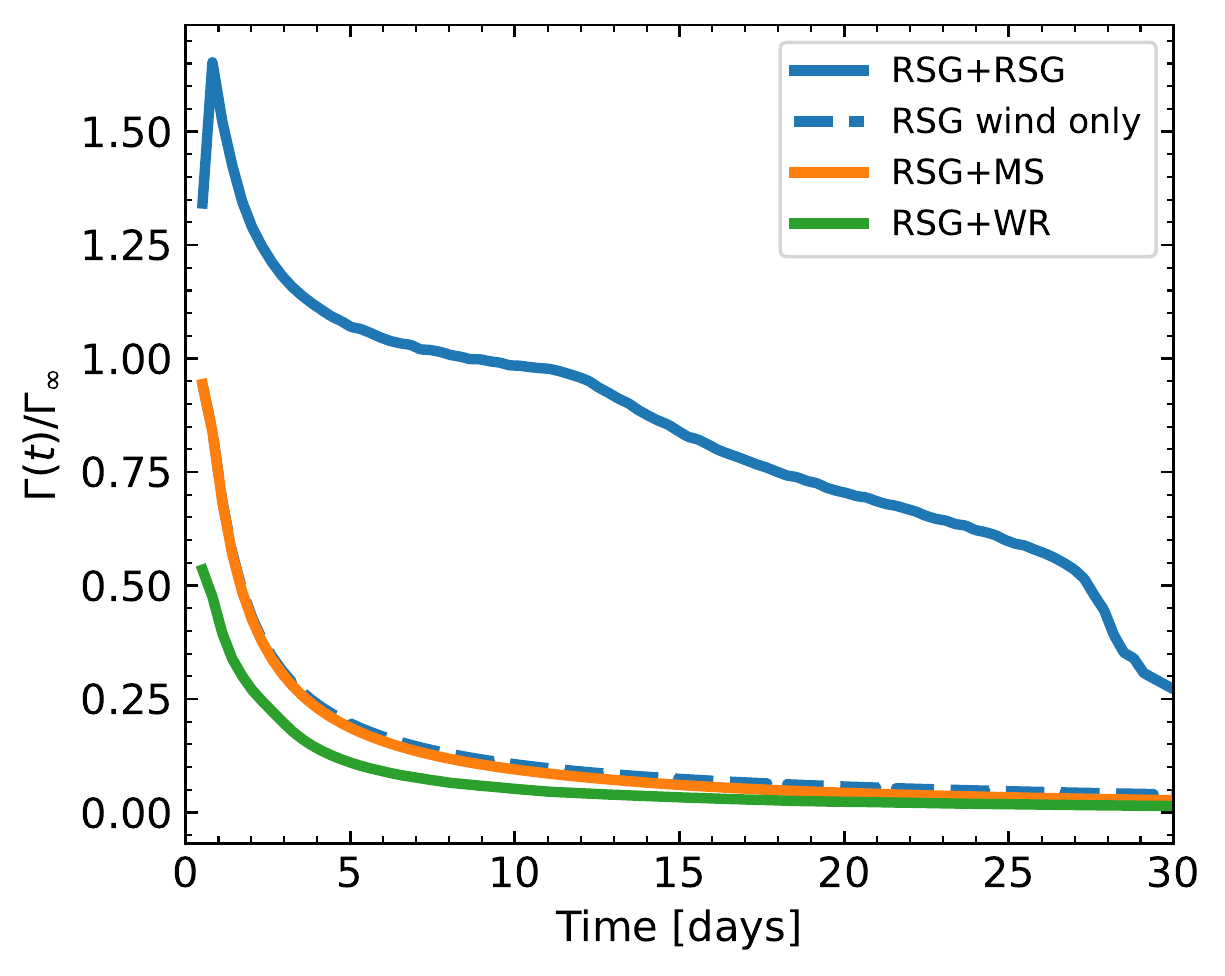}
    \caption{Angle-integrated recombination rates for different parameters. The top panel shows results for a twin binary similarly to Fig.~\ref{fig:single_equal} except that the fiducial calculation has $a=100$\,au. Bottom panel is for different combinations of stars as in Fig.~\ref{fig:single_different} with $a=100$\,au. Time delay is not taken into account.}
    \label{fig:flash_compare}
\end{figure}

In Figure~\ref{fig:flash_compare}, we show the total relative recombination rate for different binary and stellar parameters. We see that CW shells in twin binaries (top panel) consistently yield recombination signal that is by a factor of at most $10$ higher than for an unperturbed single-star wind. Most of the dependence on $\mdot$ and $v_\infty$ is absorbed in $\Gamma_\infty$, but the residual differences can be explained by $\rsh(t)$ moving slower in denser environments. Changing $a$ moves the CW shell closer or farther away from the exploding star, which affects the duration and the magnitude of the signal. With time-delay taken into account, increasing $a$ delays the rise of the recombination rate \citep{kochanek19}. The value of CW shell width $\delta_\text{cw}$ has the expected effect that is caused by the changes of the CW shell density. Increasing the wind velocities leads to a stronger relative signal, but the absolute magnitude of the effect will be significantly smaller due to $\Gamma_\infty \propto v_\infty^{-2}$.

For binaries consisting of different stars (bottom panel of Fig.~\ref{fig:flash_compare}), we see that MS and WR companions to RSG decrease the recombination signature. The reason is that the higher velocity of the companion wind increases the flow of the matter through the CW shell, which in turn decreases the density. In the case of RSG+WR binary, the recombination signature is actually even below the unperturbed single-star wind, because the fast WR wind effectively sets an outer cutoff radius for the RSG 
wind.
We do not show predictions for binaries, where the primary has a fast wind and the secondary is a RSG, which leads to very high $\Gamma/\Gamma_\infty$. These high values are caused entirely by the signal coming from the dense wind of the secondary, while $\Gamma_\infty$ is evaluated for the more tenuous primary wind.

To summarize, we find the recombination light curves due to CW shells are strongest for double RSG binaries and reach luminosities of up to $10$ times higher than for an unperturbed single-star wind with $r^{-2}$ density profile. The signal will last more than a few days for binaries with $a\gtrsim 50$\,au.

\subsection{Population estimate for non-interacting binaries}
\label{sec:pop}

\begin{figure}
    \centering
    \includegraphics[width=0.48\textwidth]{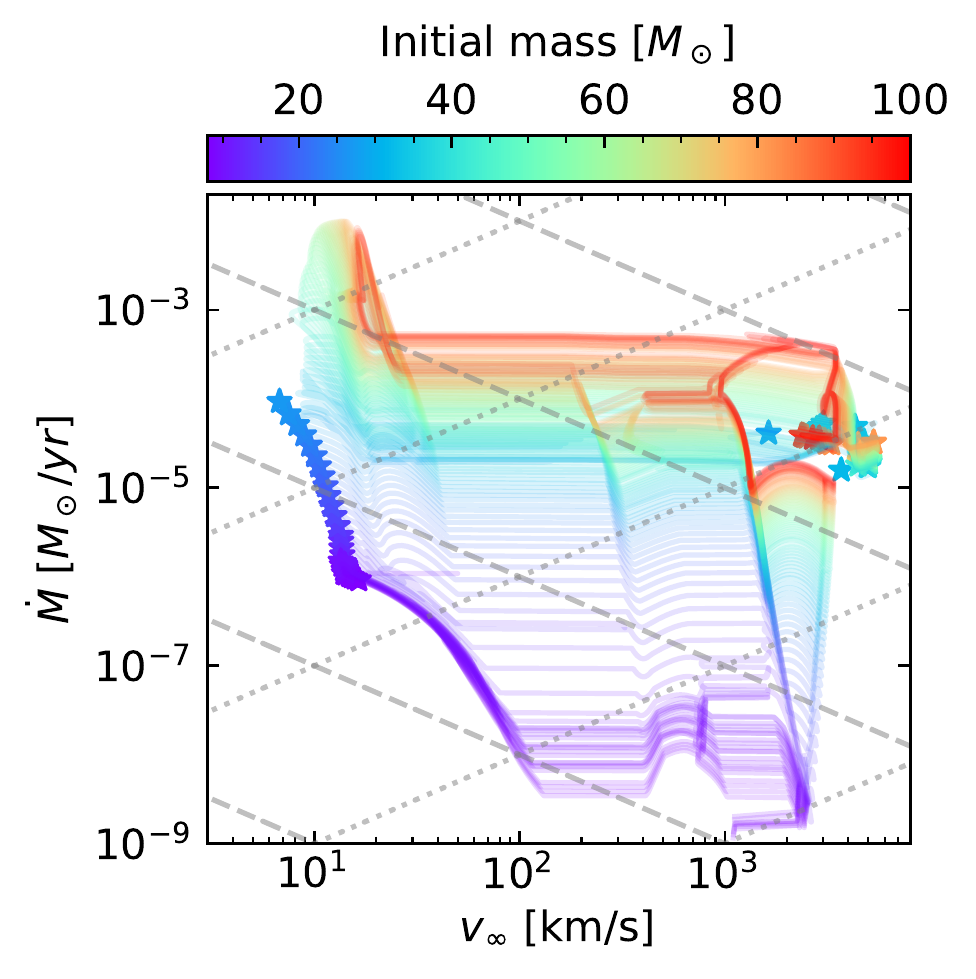}
    \caption{Wind properties of massive single stars as function of their initial mass and age based on BPASS models. Lines of different color show the evolutionary tracks. The end of the evolution is marked with stars. Lines of constant wind momentum $\mdot v_\infty$ are marked with dashed lines, lines of constant ratio $\mdot/v_\infty$ are shown with dotted lines.}
    \label{fig:bpass_single}
\end{figure}

\begin{figure}
    \centering
    \includegraphics[width=0.48\textwidth]{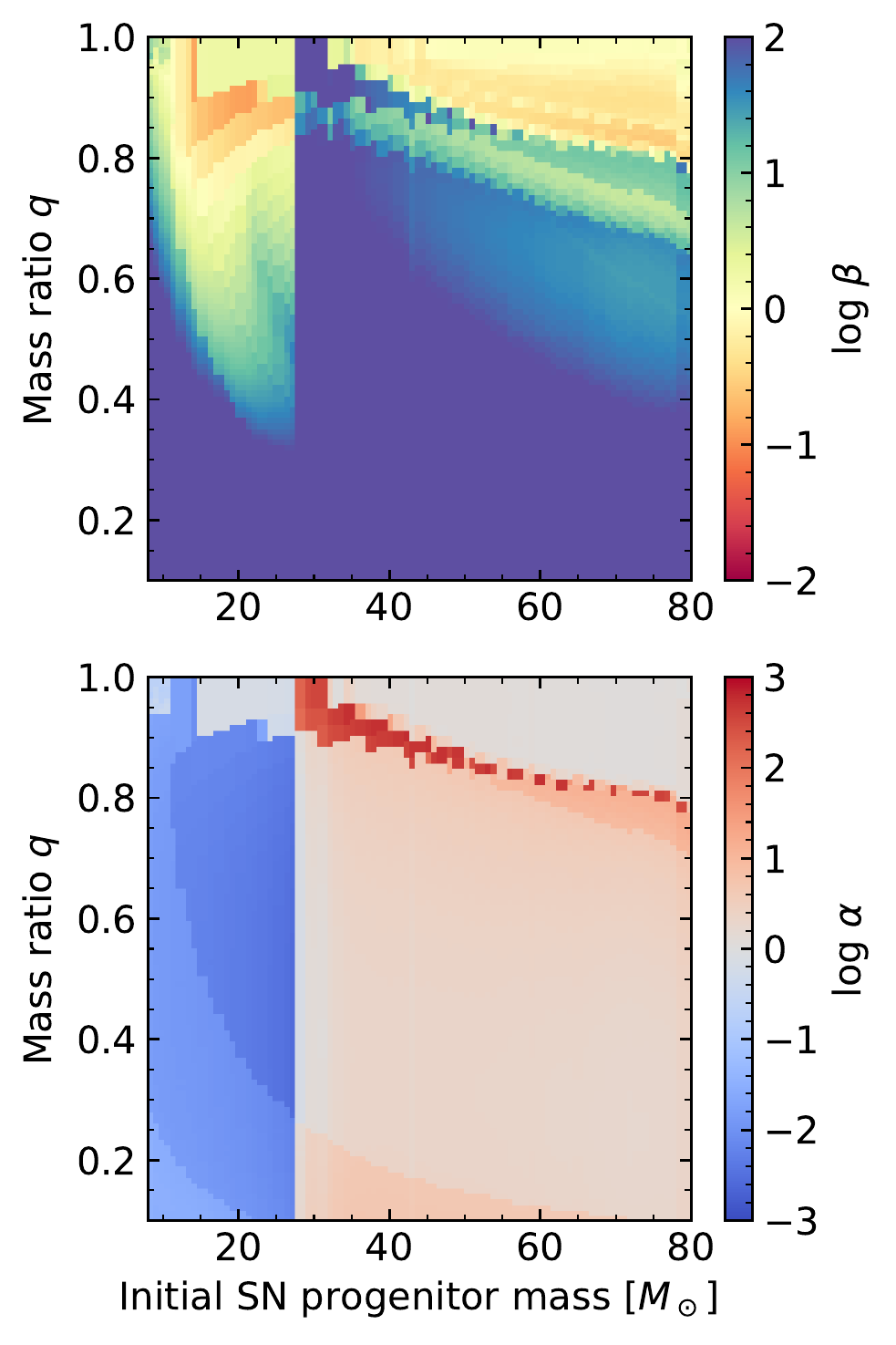}
    \caption{Colliding wind shell parameters in wide binaries based on BPASS models. Upper panel shows ratio of wind momenta $\beta$ and lower panel the ratio of wind velocities $\alpha$.}
    \label{fig:bpass_wind}
\end{figure}


Our results from previous sections imply that SN interaction with a CW shell is strongest when both stars have slow and dense winds like is the case of double RSG binaries. However, $\mdot$ required to match the luminosity of normal Type II-P SNe is about $10^{-4}\,\myr$. The signal is somewhat weaker when the companion has faster wind as is the case of MS or WR stars, and correspondingly higher $\mdot$ is required to reach the same $\Lshock$ as for the double RSG binaries. Although the higher metal content in WR winds will enhance the shock radiative efficiency, the fundamental limiting factor is still the raw shock power, which is set by the mass of the CW shell.  We estimated that signatures of CW shells in SNe occur for $a\gtrsim 50$\,au. Such long periods imply that stars in these binaries will never start interacting by mass transfer and will evolve effectively as single stars.

Here, we address the issue of how frequent shock interactions with CW shells are. We ignore complications from hierarchical triples and higher-order multiple systems and base our estimates on single-star models from \textit{Binary Population and Spectral Synthesis} version 2.2.1 \citep[BPASS][]{eldridge17,stanway18}. We choose models with metallicity $Z=0.014$. The wind mass loss rates come from \citet{dejager88} except OB and WR stars, where \citet{vink01} and \citet{nugis00} rates are used, respectively. The models were evolved until the end of core carbon burning or neon ignition and we assume that surface properties at this point are identical to the properties at the moment of core collapse. 

BPASS models provide $\mdot$, but we need to estimate also $v_\infty$. First, we calculate the Eddington factor using the results of \citet{grafener11}
\begin{equation}
    \log \Gamma = -4.813 + \log (1 + X_\text{surf}) + \log L - \log M,
    \label{eq:gamma}
\end{equation}
and calculate the escape velocity
\begin{equation}
    \vesc = \sqrt{\frac{2(1-\Gamma)GM}{R}}.
    \label{eq:vesc}
\end{equation}
We estimate the terminal wind velocity as
\begin{equation}
    \frac{v_\infty}{\vesc} = \begin{cases}
  2.6, & \text{if } \teff \ge 21 000\,\text{K}, \\
  1.3, & \text{if } 10 000\,\text{K} \le \teff < 21 000\,\text{K},\\
  0.7, & \text{if } \teff < 10000\,\text{K}\ \text{and } \log L < 3,\\
  0.7 - \frac{\log L - 3}{3.2}, & \text{if } \teff < 10000\,\text{K}\ \text{and } 3\le \log L < 4.6,\\
  0.2, & \text{if } \teff < 10000\,\text{K}\ \text{and } \log L \ge 4.6
\end{cases}
\label{eq:vinfty}
\end{equation}
which is based on the compilation of hot and cold stellar wind properties from \citet{lamers99} using the results of \citet{lamers95} and \citet{dupree87}. Equation~(\ref{eq:vinfty}) takes into account bistability jumps of hot winds and the relative decrease of terminal velocity as a function of increasing luminosity for cool luminous stars. The linear dependence on $\log L$ and the asymptotic value of $v_\infty/\vesc = 0.2$ for luminous cool stars is set to approximately match the six stars listed \citet[Table 2.3]{lamers99}.

In Figure~\ref{fig:bpass_single}, we show the evolution of $\mdot$ and $v_\infty$ for massive stars in BPASS models. Stars with initial masses $<27\,\msun$ terminate their evolution as RSG while more massive stars die as WR stars. We see that $\mdot$ approaches $10^{-4}\,\myr$ for WR stars and the most massive RSG. The models reveal that RSG and MS winds can have comparable momentum, but that the ratio $\mdot/v_\infty$, which is related to the mass in the CW shell (see Eq.~[\ref{eq:sigma}]) is significantly higher for RSG than WR or MS stars. This justifies our conclusions that CW shells in double RSG binaries have stronger observational signatures from shock-powered collisions.

We now proceed to construct binaries from single-star BPASS models. For each primary, we record the age at the end of the evolution. We interpolate in the evolutionary tracks to find stellar parameters for all secondaries with age equal to the time when the primary explodes as a SN. In Figure~\ref{fig:bpass_wind}, we show our results for the basic CW parameters $\beta$ and $\alpha$ as a function of primary mass and binary mass ratio $q$. For RSG primaries, we see that similar wind momenta $\beta \sim 1$ are found for a relatively wide range of $q \gtrsim 0.5$. For WR primaries, $\beta \approx 1$ occurs mostly for nearly identical secondaries also in the WR phase. For all other binaries, $\beta \gg 1$ because all other evolutionary phases have lower wind momenta (Fig.~\ref{fig:bpass_single}). The narrow stripe of high $\beta$ starting at $q=1$ at $M=30\,\msun$ and ending at $q=0.75$ at $M=80\,\msun$ is caused by binaries with a cool evolved secondary with a strong slow wind. The ratio of wind velocities $\alpha$ shows nearly identical pattern with one important exception: while $\beta \sim 1$ for a wide range of $q$ for RSG primaries, $\alpha \sim 1$ only for RSG secondaries, which occurs for $q>0.9$. For $a=100$\,au, cooling parameters are $\chi_A\approx 10^{-4}$ and $\sim 10$ for RSG and WR primaries, respectively. The secondaries typically have $\chi_B > 10^2$ except when they are RSGs, which occurs in a similar part of parameter space where $\alpha \gtrsim 10^2$.

We expect the strongest signatures of CW shells in double RSG binaries. For other types of binaries, the geometry of the CW shell requires so high $\mdot$ that it seems unrealistic given the usual stellar evolution models. The frequency of WR+RSG binaries will be further suppressed by the declining initial mass function and the narrow range of $q$, where this configuration occurs. As a result, we conclude that the best chances for revealing the CW shell by shock luminosity or flash ionization is for primary initial masses of $20\le M_A\le 27\,\msun$ and $q\ge 0.9$. In an optimistic scenario, the primary mass range could be expanded down to $8\,\msun$ if for some reason the RSG mass-loss rates were underestimated.

What is the frequency of such binaries among stellar population? \citet{moe17} found that O and B stars have on average $0.1$ companions with $q>0.3$ per decade of $P$ for $\log P \gtrsim 4$, which implies that roughly $0.22$ of all massive stars have a $q>0.3$ companion at $50 \lesssim a \lesssim 1500$\,au, where the effect of CW shells on SNe is observable. \citet{moe17} did not find any significant twin fraction for massive stars with such long $P$. Since the companion is so distant, its mass function is very similar to an independent draw from the initial mass function. \citet{moe17} estimated that for these separations the power-law exponent describing the distribution of $q$ is about $-2.0$, which implies that the fraction of binaries with $q>0.9$ is $0.048$. Consequently, the fraction of massive stars with the companion of the right mass and separation to give observable effect of CW is about $1.0$~per~cent. Finally, if we require that only primaries with $20\le M\le 27\,\msun$ have sufficiently strong winds and assuming \citet{salpeter55} initial mass function, the fraction drops by another factor of $0.1$ to a total of $0.1$~per~cent. Our estimate implies that CW shells can explain only a small number of individual peculiar events, but do not systematically influence SN population.

\section{Discussions and conclusion}
\label{sec:disc}

In this paper, we explored SN explosions interacting with a CW shell in a binary star system and their consequences for shock-powered light curves and flash ionization signatures. In Section~\ref{sec:csm}, we calibrated the analytic model of CW shells of \citet{canto96} using a suite of adaptive mesh refinement hydrodynamical simulations of \citet{calderon20a,calderon20b} covering both adiabatic and radiatively-unstable regimes. We heuristically included orbital motion and constructed the final semi-analytic model of the CSM distribution including stellar winds of both components and the CW shell. In Section~\ref{sec:lc}, we generalized the thin-shell dynamics model of \citet{metzger14} to angularly-dependent CSM distributions.

We calculate (Section~\ref{sec:simple}) that the highest shock luminosity occurs when the winds of both stars are dense such as is the case in double RSG binaries. We find that $\mdot \gtrsim 10^{-4}\,\myr$ is required to exceed optical luminosities of normal Type IIP SNe. We estimate that CW shells are best detectable for $50 \lesssim a \lesssim 1500$\,au. For smaller separations, the shock interactions is weaker and occurs too early in the light curve to be distinguishable from the SN shock cooling emission. For larger separations, the shock luminosity rises too late to influence the optically-thick part of the SN light curves. We find that the flash ionization signature (Section~\ref{sec:flash}) is also strongest for double RSG binaries, because companions with faster wind significantly reduce the mass in the shell. Binary separation most significantly affects the time delay of the recombination signal due to light travel time effects, which we do not explicitly model here, and the duration of the signal. By considering the statistics of binaries (Section~\ref{sec:pop}), we estimate that at most $1$~per~cent of all SNe will show CW shell signatures in the first $\sim 300$\,days, but a more realistic estimate taking into account mass-loss rates as a function of primary's mass is at least factor of $10$ smaller.

There are three significant effects that we did not include and which might modify our results. First, stellar winds accelerate to their asymptotic velocity over spatial scales, which might be comparable to the binary separation. The exponent parameterizing radial dependence of in the common velocity law ranges from about $0.5$--$1$ for hot stars to about $1.5$--$3.5$ in the case of RSGs \citep[e.g.][]{baade96}. This effect is important for early SN light curves \citep{moriya17,moriya18}. We cannot self-consistently implement wind acceleration in our model, because the model of the CW shell would cease to be analytic. However, we can get a crude estimate by replacing the wind component of the CSM distribution by an accelerated wind. We find very little difference in shock luminosity with respect to the fiducial model (bright blue line in Fig.~\ref{fig:single_equal}). We also estimated time evolution of recombination flux for single-star accelerating wind (blue dashed line in Fig.~\ref{fig:gamma_angle}). We see that the light curve peaks at much higher values, which would lead to an inference of much higher $\Gamma_\infty$ if interpreted with the simple $\rho \propto r^{-2}$ wind model. Because the densities near the star are much higher, the wind acceleration signal is much stronger than the recombination from the CW shell. Furthermore, since the high densities are located near the star, the time delay would be minimal, which would pose further difficulties in explaining events like SN2013fs \citep{yaron17,kochanek19}.

Second, one way to increase the signal from CW shell interaction would be if the primary mass-loss rate increased above the values given by the prescriptions used in stellar evolution codes shortly before core collapse. The wind of the companion could compress this mass ejection or enhanced wind similarly to what happens in a CW shell. Although, it is now possible to estimate the time-scale and amount of ejected mass as a function of $M$ in the context of wave-driven mass-loss \citep[e.g.][]{wu21,leung21} and the first 
pre-SN brightening of a Type IIP/IIL progenitor has now been detected \citep{johnson18,jacobsongalan21}, the understanding of such events remains insufficient. Nonetheless, the CW-like shell that might be formed as a result could be interpreted with the theory developed here.

Third, the relatively low frequency of observable SN shock interactions with a CW shell is mainly caused by the requirement of nearly equal-mass companions, which are rare among long-period binaries. The frequency  might be influenced by one of the stars actually being a close binary, which is expected for massive stars \citep{moe17}. It is not clear whether binary interactions like mergers or rejuvenation in the close component would improve or reduce the timing chances of having a double RSG binary at the moment of the first SN explosion.

Are there physical effects due to the binary companion other than colliding winds  that could compress the wind of the primary? \citet{kochanek19} considered perturbations to the primary wind from the photoionizing flux of the secondary. However, luminosity and ionizing flux also steeply depend on companion's mass and this physical effect likely faces the same rate problem as the colliding winds. Another option is a simple gravitational focusing of the primary wind by the companion and a formation of a spiral wake, where the gas could radiatively cool to high densities. There are simulations of this process mostly in the context of wind Roche-lobe overflow of red giants and AGB stars \citep[e.g.][]{mohamed07,devalborro17,saladino19,chen20,schroder21}. Gravitational focusing would work best for stars with slow winds and wide binaries, where the wind and orbital velocities are comparable. However, gravitational focusing only affects wind in the orbital plane of the binary and the resulting overdensity will necessarily cover only a small fraction of the solid angle. It is not clear, whether this type of overdensity would produce a stronger flash ionization signal than the high-density wind acceleration zone near the star or the extended structure recently identified in 3D simulations of RSG envelopes \citep{goldberg21}.

\section*{Acknowledgements}
OP appreciates discussions with Chris Kochanek about flash ionization. The research of OP and DC has been supported by Horizon 2020 ERC Starting Grant `Cat-In-hAT' (grant agreement no. 803158).
Most of the calculations and visualizations in this work were performed with \textsc{matplotlib} \citep{hunter07}, \textsc{scipy} \citep{virtanen20}, and \textsc{numpy} \citep{harris20}. The analysis of the AMR simulations was carried out making use of the package \textsc{yt} \citep{turk11}.

\section*{Data availability}
The data underlying this article will be shared on reasonable request to the corresponding author.

\bibliographystyle{mnras}
\bibliography{mybib}

\bsp	
\label{lastpage}
\end{document}